\documentclass[longauth,traditabstract]{aa}
\usepackage{graphicx}
\usepackage{amsmath,amsfonts,amssymb}
\usepackage{bm}
\usepackage{txfonts}

\usepackage[citecolor=blue,pdfa=true,urlcolor=magenta,colorlinks=True]{hyperref}

\usepackage{color}
\usepackage{fixltx2e}
\usepackage{natbib}

\usepackage{multirow}
\usepackage{epsf}
\usepackage{epsfig}
\usepackage{appendix}

\usepackage{ifthen}

\usepackage[T1]{fontenc}
\usepackage{lmodern}
\usepackage{ifxetex,ifluatex}

\usepackage{dblfloatfix}
\usepackage{morefloats}
\usepackage{booktabs}
\usepackage[switch]{lineno}
\usepackage{tabularx}
\bibpunct{(}{)}{;}{a}{}{,}
\usepackage[bottom]{footmisc}
\def\setsymbol#1#2{\expandafter\def\csname #1\endcsname{#2}}
\def\getsymbol#1{\csname #1\endcsname}

\def\Planck{\textit{Planck}}

\newbox\tablebox    \newdimen\tablewidth
\def\leaderfil{\leaders\hbox to 5pt{\hss.\hss}\hfil}
\def\endPlancktable{\tablewidth=\columnwidth 
    $$\hss\copy\tablebox\hss$$
    \vskip-\lastskip\vskip -2pt}

\def\tablenote#1 #2\par{\begingroup \parindent=0.8em
    \abovedisplayshortskip=0pt\belowdisplayshortskip=0pt
    \noindent
    $$\hss\vbox{\hsize\tablewidth \hangindent=\parindent \hangafter=1 \noindent
    \hbox to \parindent{$^#1$\hss}\strut#2\strut\par}\hss$$
    \endgroup}
\def\doubleline{\vskip 3pt\hrule \vskip 1.5pt \hrule \vskip 5pt}

\def\L2{\ifmmode L_2\else $L_2$\fi}

\def\DeltaT{\ifmmode \Delta T\else $\Delta T$\fi}
\def\deltat{\ifmmode \Delta t\else $\Delta t$\fi}
\def\fknee{\ifmmode f_{\rm knee}\else $f_{\rm knee}$\fi}
\def\Fmax{\ifmmode F_{\rm max}\else $F_{\rm max}$\fi}
\def\solar{\ifmmode{\rm M}_{\mathord\odot}\else${\rm M}_{\mathord\odot}$\fi}
\def\Msolar{\ifmmode{\rm M}_{\mathord\odot}\else${\rm M}_{\mathord\odot}$\fi}
\def\Lsolar{\ifmmode{\rm L}_{\mathord\odot}\else${\rm L}_{\mathord\odot}$\fi}
\def\inv{\ifmmode^{-1}\else$^{-1}$\fi}
\def\mo{\ifmmode^{-1}\else$^{-1}$\fi}
\def\sup#1{\ifmmode ^{\rm #1}\else $^{\rm #1}$\fi}
\def\expo#1{\ifmmode \times 10^{#1}\else $\times 10^{#1}$\fi}
\def\,{\thinspace}
\def\lsim{\mathrel{\raise .4ex\hbox{\rlap{$<$}\lower 1.2ex\hbox{$\sim$}}}}
\def\gsim{\mathrel{\raise .4ex\hbox{\rlap{$>$}\lower 1.2ex\hbox{$\sim$}}}}

\def\simprop{\mathrel{\raise .4ex\hbox{\rlap{$\propto$}\lower 1.2ex\hbox{$\sim$}}}}
\def\deg{\ifmmode^\circ\else$^\circ$\fi}
\def\pdeg{\ifmmode $\setbox0=\hbox{$^{\circ}$}\rlap{\hskip.11\wd0 .}$^{\circ}
          \else \setbox0=\hbox{$^{\circ}$}\rlap{\hskip.11\wd0 .}$^{\circ}$\fi}
\def\arcs{\ifmmode {^{\scriptstyle\prime\prime}}
          \else $^{\scriptstyle\prime\prime}$\fi}
\def\arcm{\ifmmode {^{\scriptstyle\prime}}
          \else $^{\scriptstyle\prime}$\fi}
\newdimen\sa  \newdimen\sb
\def\parcs{\sa=.07em \sb=.03em
     \ifmmode \hbox{\rlap{.}}^{\scriptstyle\prime\kern -\sb\prime}\hbox{\kern -\sa}
     \else \rlap{.}$^{\scriptstyle\prime\kern -\sb\prime}$\kern -\sa\fi}
\def\parcm{\sa=.08em \sb=.03em
     \ifmmode \hbox{\rlap{.}\kern\sa}^{\scriptstyle\prime}\hbox{\kern-\sb}
     \else \rlap{.}\kern\sa$^{\scriptstyle\prime}$\kern-\sb\fi}
\def\ra[#1 #2 #3.#4]{#1\sup{h}#2\sup{m}#3\sup{s}\llap.#4}
\def\dec[#1 #2 #3.#4]{#1\deg#2\arcm#3\arcs\llap.#4}
\def\deco[#1 #2 #3]{#1\deg#2\arcm#3\arcs}
\def\rra[#1 #2]{#1\sup{h}#2\sup{m}}

\def\dots{\relax\ifmmode \ldots\else $\ldots$\fi}
\def\WHzsr{\ifmmode $W\,Hz\mo\,sr\mo$\else W\,Hz\mo\,sr\mo\fi}
\def\mHz{\ifmmode $\,mHz$\else \,mHz\fi}
\def\GHz{\ifmmode $\,GHz$\else \,GHz\fi}
\def\mKs{\ifmmode $\,mK\,s$^{1/2}\else \,mK\,s$^{1/2}$\fi}
\def\muKs{\ifmmode \,\mu$K\,s$^{1/2}\else \,$\mu$K\,s$^{1/2}$\fi}
\def\muKRJs{\ifmmode \,\mu$K$_{\rm RJ}$\,s$^{1/2}\else \,$\mu$K$_{\rm RJ}$\,s$^{1/2}$\fi}
\def\muKHz{\ifmmode \,\mu$K\,Hz$^{-1/2}\else \,$\mu$K\,Hz$^{-1/2}$\fi}
\def\MJysr{\ifmmode \,$MJy\,sr\mo$\else \,MJy\,sr\mo\fi}
\def\MJysrmK{\ifmmode \,$MJy\,sr\mo$\,mK$_{\rm CMB}\mo\else \,MJy\,sr\mo\,mK$_{\rm CMB}\mo$\fi}
\def\microns{\ifmmode \,\mu$m$\else \,$\mu$m\fi}

\def\muK{\ifmmode \,\mu$K$\else \,$\mu$\hbox{K}\fi}
\def\microK{\ifmmode \,\mu$K$\else \,$\mu$\hbox{K}\fi}
\def\muW{\ifmmode \,\mu$W$\else \,$\mu$\hbox{W}\fi}
\def\kms{\ifmmode $\,km\,s$^{-1}\else \,km\,s$^{-1}$\fi}
\def\kmsMpc{\ifmmode $\,\kms\,Mpc\mo$\else \,\kms\,Mpc\mo\fi}
\providecommand{\sorthelp}[1]{}

\def\fig{\frenchspacing Fig. }

\def\eqref#1{(\ref{#1})}

\def\smicano{{\tt SMICA-NOSZ}}
\def\nilc{{\tt NILC}}
\def\milca{{\tt MILCA}}
\def\twodilc{{\tt 2D-ILC}}

\def\healpix{\texttt{HEALPix}}

\def\nside{$N_{\mathrm{side}}$}

\def\curl{\mathcal}
\def\({\left(}
\def\){\right)}

\def\and{\quad \mbox{and} \quad}
\def\barQ{\kern2pt\overline{\kern-2pt\curl{Q}}}

\def\barR{\kern2pt\overline{\kern-2pt\curl{R}}}

\def\bargamma{\kern2pt\overline{\kern-2pt\gamma}}

\def\leaderfil{\leaders\hbox to 5pt{\hss.\hss}\hfil}

\newcommand{\be}{\begin{equation}}
\newcommand{\ee}{\end{equation}}

\renewcommand{\L}[0]{\mathbf{L}}

\def\inv{^{-1}}

\newcommand{\spinup}{\;\raise1.0pt\hbox{$'$}\hskip-6pt\partial\;}
\newcommand{\spindown}{\;\overline{\raise1.0pt\hbox{$'$}\hskip-6pt\partial}\;}

\graphicspath{{./figures/}}

\makeatletter
\newcommand*{\StrutboxHeight}{\ht\strutbox@}
\newcommand*{\StrutboxDepth}{\dp\strutbox@}
\makeatother

\let\vec\mathbf

\defcitealias{planck2013-p09}{PCIS13}

\title{\Planck\ intermediate results. LVI. Detection of the CMB dipole through
modulation of the thermal Sunyaev-Zeldovich effect: Eppur si muove
II}

\author{\small
Planck Collaboration: Y.~Akrami\inst{11, 46, 47}
\and
M.~Ashdown\inst{53, 4}
\and
J.~Aumont\inst{75}
\and
C.~Baccigalupi\inst{62}
\and
M.~Ballardini\inst{16, 32}
\and
A.~J.~Banday\inst{75, 7}
\and
R.~B.~Barreiro\inst{49}
\and
N.~Bartolo\inst{20, 50}
\and
S.~Basak\inst{67}
\and
K.~Benabed\inst{45, 69}
% ...
}
\institute{\small
AIM, CEA, CNRS, Universit\'{e} Paris-Saclay, Universit\'{e} Paris-Diderot, Sorbonne Paris Cit\'{e}, F-91191 Gif-sur-Yvette, France\goodbreak
\and
APC, AstroParticule et Cosmologie, Universit\'{e} Paris Diderot, CNRS/IN2P3, CEA/lrfu, Observatoire de Paris, Sorbonne Paris Cit\'{e}, 10, rue Alice Domon et L\'{e}onie Duquet, 75205 Paris Cedex 13, France\goodbreak
\and
African Institute for Mathematical Sciences, 6-8 Melrose Road, Muizenberg, Cape Town, South Africa\goodbreak
\and
Astrophysics Group, Cavendish Laboratory, University of Cambridge, J J Thomson Avenue, Cambridge CB3 0HE, U.K.\goodbreak
\and
Astrophysics \& Cosmology Research Unit, School of Mathematics, Statistics \& Computer Science, University of KwaZulu-Natal, Westville Campus, Private Bag X54001, Durban 4000, South Africa\goodbreak
\and
CITA, University of Toronto, 60 St. George St., Toronto, ON M5S 3H8, Canada\goodbreak
\and
CNRS, IRAP, 9 Av. colonel Roche, BP 44346, F-31028 Toulouse cedex 4, France\goodbreak
\and
California Institute of Technology, Pasadena, California, U.S.A.\goodbreak
\and
Computational Cosmology Center, Lawrence Berkeley National Laboratory, Berkeley, California, U.S.A.\goodbreak
\and
D\'{e}partement de Physique Th\'{e}orique, Universit\'{e} de Gen\`{e}ve, 24, Quai E. Ansermet,1211 Gen\`{e}ve 4, Switzerland\goodbreak
\and
D\'{e}partement de Physique, \'{E}cole normale sup\'{e}rieure, PSL Research University, CNRS, 24 rue Lhomond, 75005 Paris, France\goodbreak
\and
Departamento de Astrof\'{i}sica, Universidad de La Laguna (ULL), E-38206 La Laguna, Tenerife, Spain\goodbreak
\and
Departamento de F\'{\i}sica, Universidad de Oviedo, C/ Federico Garc\'{\i}a Lorca, 18 , Oviedo, Spain\goodbreak
\and
Department of Astrophysics/IMAPP, Radboud University, P.O. Box 9010, 6500 GL Nijmegen, The Netherlands\goodbreak
\and
Department of Physics \& Astronomy, University of British Columbia, 6224 Agricultural Road, Vancouver, British Columbia, Canada\goodbreak
\and
Department of Physics \& Astronomy, University of the Western Cape, Cape Town 7535, South Africa\goodbreak
\and
Department of Physics and Astronomy, York University, Toronto, Ontario, M3J 1P3, Canada\goodbreak
\and
Department of Physics, Gustaf H\"{a}llstr\"{o}min katu 2a, University of Helsinki, Helsinki, Finland\goodbreak
\and
Department of Physics, Princeton University, Princeton, New Jersey, U.S.A.\goodbreak
\and
Dipartimento di Fisica e Astronomia G. Galilei, Universit\`{a} degli Studi di Padova, via Marzolo 8, 35131 Padova, Italy\goodbreak
\and
Dipartimento di Fisica e Scienze della Terra, Universit\`{a} di Ferrara, Via Saragat 1, 44122 Ferrara, Italy\goodbreak
\and
Dipartimento di Fisica, Universit\`{a} La Sapienza, P. le A. Moro 2, Roma, Italy\goodbreak
\and
Dipartimento di Fisica, Universit\`{a} degli Studi di Milano, Via Celoria, 16, Milano, Italy\goodbreak
\and
Dipartimento di Fisica, Universit\`{a} degli Studi di Trieste, via A. Valerio 2, Trieste, Italy\goodbreak
\and
Dipartimento di Fisica, Universit\`{a} di Roma Tor Vergata, Via della Ricerca Scientifica, 1, Roma, Italy\goodbreak
\and
European Space Agency, ESAC, Planck Science Office, Camino bajo del Castillo, s/n, Urbanizaci\'{o}n Villafranca del Castillo, Villanueva de la Ca\~{n}ada, Madrid, Spain\goodbreak
\and
European Space Agency, ESTEC, Keplerlaan 1, 2201 AZ Noordwijk, The Netherlands\goodbreak
\and
Gran Sasso Science Institute, INFN, viale F. Crispi 7, 67100 L'Aquila, Italy\goodbreak
\and
HEP Division, Argonne National Laboratory, Lemont, IL 60439, USA\goodbreak
\and
Helsinki Institute of Physics, Gustaf H\"{a}llstr\"{o}min katu 2, University of Helsinki, Helsinki, Finland\goodbreak
\and
IFPU - Institute for Fundamental Physics of the Universe, Via Beirut 2, 34014 Trieste, Italy\goodbreak
\and
INAF - OAS Bologna, Istituto Nazionale di Astrofisica - Osservatorio di Astrofisica e Scienza dello Spazio di Bologna, Area della Ricerca del CNR, Via Gobetti 101, 40129, Bologna, Italy\goodbreak
\and
INAF - Osservatorio Astronomico di Padova, Vicolo dell'Osservatorio 5, Padova, Italy\goodbreak
\and
INAF - Osservatorio Astronomico di Trieste, Via G.B. Tiepolo 11, Trieste, Italy\goodbreak
\and
INAF - Osservatorio Astronomico di Trieste, via G. B. Tiepolo 11, I-34143 Trieste, Italy\goodbreak
\and
INAF, Istituto di Radioastronomia, Via Piero Gobetti 101, I-40129 Bologna, Italy\goodbreak
\and
INAF/IASF Milano, Via E. Bassini 15, Milano, Italy\goodbreak
\and
INFN - CNAF, viale Berti Pichat 6/2, 40127 Bologna, Italy\goodbreak
\and
INFN, Sezione di Bologna, viale Berti Pichat 6/2, 40127 Bologna, Italy\goodbreak
\and
INFN, Sezione di Ferrara, Via Saragat 1, 44122 Ferrara, Italy\goodbreak
\and
INFN, Sezione di Milano, Via Celoria 16, Milano, Italy\goodbreak
\and
INFN, Sezione di Roma 2, Universit\`{a} di Roma Tor Vergata, Via della Ricerca Scientifica, 1, Roma, Italy\goodbreak
\and
Imperial College London, Astrophysics group, Blackett Laboratory, Prince Consort Road, London, SW7 2AZ, U.K.\goodbreak
\and
Institut d'Astrophysique Spatiale, CNRS, Univ. Paris-Sud, Universit\'{e} Paris-Saclay, B\^{a}t. 121, 91405 Orsay cedex, France\goodbreak
\and
Institut d'Astrophysique de Paris, CNRS (UMR7095), 98 bis Boulevard Arago, F-75014, Paris, France\goodbreak
\and
Institute Lorentz, Leiden University, PO Box 9506, Leiden 2300 RA, The Netherlands\goodbreak
\and
Institute of Theoretical Astrophysics, University of Oslo, Blindern, Oslo, Norway\goodbreak
\and
Instituto de Astrof\'{\i}sica de Canarias, C/V\'{\i}a L\'{a}ctea s/n, La Laguna, Tenerife, Spain\goodbreak
\and
Instituto de F\'{\i}sica de Cantabria (CSIC-Universidad de Cantabria), Avda. de los Castros s/n, Santander, Spain\goodbreak
\and
Istituto Nazionale di Fisica Nucleare, Sezione di Padova, via Marzolo 8, I-35131 Padova, Italy\goodbreak
\and
Jet Propulsion Laboratory, California Institute of Technology, 4800 Oak Grove Drive, Pasadena, California, U.S.A.\goodbreak
\and
Jodrell Bank Centre for Astrophysics, Alan Turing Building, School of Physics and Astronomy, The University of Manchester, Oxford Road, Manchester, M13 9PL, U.K.\goodbreak
\and
Kavli Institute for Cosmology Cambridge, Madingley Road, Cambridge, CB3 0HA, U.K.\goodbreak
\and
Laboratoire de Physique Subatomique et Cosmologie, Universit\'{e} Grenoble-Alpes, CNRS/IN2P3, 53, rue des Martyrs, 38026 Grenoble Cedex, France\goodbreak
\and
Laboratoire de Physique Th\'{e}orique, Universit\'{e} Paris-Sud 11 \& CNRS, B\^{a}timent 210, 91405 Orsay, France\goodbreak
\and
Lawrence Berkeley National Laboratory, Berkeley, California, U.S.A.\goodbreak
\and
Low Temperature Laboratory, Department of Applied Physics, Aalto University, Espoo, FI-00076 AALTO, Finland\goodbreak
\and
Max-Planck-Institut f\"{u}r Astrophysik, Karl-Schwarzschild-Str. 1, 85741 Garching, Germany\goodbreak
\and
Mullard Space Science Laboratory, University College London, Surrey RH5 6NT, U.K.\goodbreak
\and
National Centre for Nuclear Research, ul. L. Pasteura 7, 02-093 Warsaw, Poland\goodbreak
\and
Perimeter Institute for Theoretical Physics, Waterloo ON N2L 2Y5, Canada\goodbreak
\and
SISSA, Astrophysics Sector, via Bonomea 265, 34136, Trieste, Italy\goodbreak
\and
San Diego Supercomputer Center, University of California, San Diego, 9500 Gilman Drive, La Jolla, CA 92093, USA\goodbreak
\and
School of Physics and Astronomy, Cardiff University, Queens Buildings, The Parade, Cardiff, CF24 3AA, U.K.\goodbreak
\and
School of Physics and Astronomy, Sun Yat-sen University, 2 Daxue Rd, Tangjia, Zhuhai, China\goodbreak
\and
School of Physics and Astronomy, University of Nottingham, Nottingham NG7 2RD, U.K.\goodbreak
\and
School of Physics, Indian Institute of Science Education and Research Thiruvananthapuram, Maruthamala PO, Vithura, Thiruvananthapuram 695551, Kerala, India\goodbreak
\and
Simon Fraser University, Department of Physics, 8888 University Drive, Burnaby BC, Canada\goodbreak
\and
Sorbonne Universit\'{e}, CNRS, UMR 7095, Institut d'Astrophysique de Paris, 98 bis bd Arago, 75014 Paris, France\goodbreak
\and
Sorbonne Universit\'{e}, Observatoire de Paris, Universit\'{e} PSL, \'{E}cole normale sup\'{e}rieure, CNRS, LERMA, F-75005, Paris, France\goodbreak
\and
Space Research Institute (IKI), Russian Academy of Sciences, Profsoyuznaya Str, 84/32, Moscow, 117997, Russia\goodbreak
\and
Space Science Data Center - Agenzia Spaziale Italiana, Via del Politecnico snc, 00133, Roma, Italy\goodbreak
\and
Space Sciences Laboratory, University of California, Berkeley, California, U.S.A.\goodbreak
\and
The Oskar Klein Centre for Cosmoparticle Physics, Department of Physics, Stockholm University, AlbaNova, SE-106 91 Stockholm, Sweden\goodbreak
\and
Universit\'{e} de Toulouse, UPS-OMP, IRAP, F-31028 Toulouse cedex 4, France\goodbreak
% ...
}

\authorrunning{Planck Collaboration}
\titlerunning{Eppur si muove II}

\begin{document}

\abstract{The largest temperature anisotropy in the cosmic microwave background (CMB) is the dipole, which has been measured with increasing accuracy for more than three decades, particularly with the \Planck\ satellite. The simplest interpretation of the dipole is that it is due to our motion with respect to the rest frame of the CMB. Since current CMB experiments infer temperature anisotropies from angular intensity variations, the dipole modulates the temperature anisotropies with the same frequency dependence as the thermal Sunyaev-Zeldovich (tSZ) effect. We present the first, and significant, detection of this signal in the tSZ maps and find that it is consistent with direct measurements of the CMB dipole, as expected. The signal contributes power in the tSZ maps, which is modulated in a quadrupolar pattern, and we estimate its contribution to the tSZ bispectrum, noting that it contributes negligible noise to the bispectrum at relevant scales.}


\maketitle

\section{Introduction}
\label{sec:intro}
In the study of cosmic microwave background (CMB) anisotropies, the largest signal is the dipole. This is mainly due to our local motion with respect to the CMB rest frame and it has been
previously measured in \citet{Kogut1993}, \citet{Fixsen1996}, and \citet{hinshaw2009}, and most recently in \citet{planck2016-l01}, \citet{planck2016-l02}, and
\citet{planck2016-l03}. Taking the
large dipole as being solely caused by our motion, the velocity is $v = (369.82 \pm
0.11) \kms$ in the direction $(l, b) = (264\pdeg021 \pm 0\pdeg011, 48\pdeg253 \pm
0\pdeg005)$ \citep{planck2016-l01}.
A velocity boost has secondary effects, such as aberration and a frequency-dependent dipolar-modulation of the CMB anisotropies \citep{Challinor2002, Burles2006}. These two effects were first measured using \Planck\footnote{\Planck\ (\url{http://www.esa.int/Planck}) is a project of the European Space Agency (ESA) with instruments provided by two scientific consortia funded by ESA member states and led by Principal Investigators from France and Italy, telescope reflectors provided through a collaboration between ESA and a scientific consortium led and funded by Denmark, and additional contributions from NASA (USA).} data, as described in \citet{planck2013-pipaberration}.
The frequency-dependent part of the dipolar-modulation signal, however, is agnostic to the source of the large CMB dipole. Therefore, its measurement is an independent determination of the CMB dipole. While it may be tempting to use this measure to detect an intrinsic dipole, it has been shown that an intrinsic dipole and a dipole induced by a velocity boost would have the same dipolar-modulation signature on the sky \citep{Challinor2002, Notari2015}.

In \citet{Notari2015}, it was pointed out that the frequency dependence
of the dipolar modulation signal could be exploited to achieve a detection with
stronger significance than that of \citet{planck2013-pipaberration}. The signal
comes from a frequency derivative of the CMB anisotropies' frequency function
and, thus, has essentially the same frequency dependence as the thermal
Sunyaev-Zeldovich (tSZ) effect. 
The dipole-induced quadropule, or "kinematic quadrupole", would also have the same frequency dependence as the tSZ effect \citep{kamionkowski2003}; however, as the quadrupole is less well constrained, a significant detection was not made in this study. 
Therefore a map of the tSZ effect must contain
a copy of the dipole-modulated CMB anisotropies, so an appropriate cross-correlation of
the CMB anisotropies with the tSZ effect would be able to pull out the signal.
In principle, it could also contribute a bias and source of noise in the bispectrum
of the tSZ effect. This is potentially important because the tSZ effect is highly
non-Gaussian and much of its information content lies in the bispectrum
\citep{Rubino-Martin2003, Bhattacharya2012, planck2013-p05b,planck2014-a28}.

In this paper we further investigate the CMB under a boost,
including tSZ effects \citep{Chluba2005,Notari2015}. We explicitly measure
the dipole using a harmonic-space-based method, similar to that outlined in
\citet{Notari2015}, and using a new map-space-based analysis,
with consistent results. We also estimate the contamination in
the tSZ bispectrum, finding that it is a negligible source of noise.

The structure of the paper is as follows.  We describe the nature of the signal
we are looking for in Sect.~\ref{sec:signal}.  The data that we use, including
the choice of CMB maps, tSZ maps, and masks, are described in
Sect.~\ref{sec:data}.  The analysis is presented in Sect.~\ref{sec:analysis},
and the results in Sect.~\ref{sec:results}, separately for the multipole-based
and map-based methods.  We briefly discuss some potential systematic effects
in Sect.~\ref{sec:systematics} and we conclude in Sect.~\ref{sec:conclusions}.
We discuss the issues related to the
tSZ bispectrum in Appendix~\ref{sec:bispectrum}, the results coming from the use of an
alternative tSZ map in Appendix~\ref{app:nilc}, and how the results could
become stronger if we used a less conservative multipole cut in
Appendix~\ref{app:higherL}.

\section{Signal}
\label{sec:signal}
Here we derive the signal we are looking for. First let us introduce some useful
definitions:
\begin{align}
  x &\equiv \frac{h\nu}{k_\mathrm{B} T}; \\
  I &\equiv \frac{2k_\mathrm{B}^3T^3}{h^2c^2} \frac{x^3}{e^x - 1}; \\
  f(x) &\equiv \frac{xe^x}{e^x - 1}; \\
  Y(x) &\equiv x \frac{e^x + 1}{e^x - 1} - 4.
  \label{eq:definitions}
\end{align}
These are the dimensionless frequency, the Planck blackbody intensity function,
the frequency dependence of the CMB anisotropies, and the relative frequency
dependence of the tSZ effect, respectively. Here $h$ is Planck's constant, $k_\mathrm{B}$ is the Boltzmann constant, and $c$ is the speed of light.

To first order, anisotropies of intensity take the form
\begin{align}
  \delta I(\vec{\hat{n}}) &= If(x) \left[\frac{\delta T(\vec{\hat{n}})}{T_0} + y(\vec{\hat{n}}) Y(x) \right],
  \label{eq:firstorder}
\end{align}
where the first term represents the CMB anisotropies,\footnote{We note that for
brevity we have not written the kinetic Sunyaev-Zeldovich (kSZ) effect; however,
its presence is accounted for in our analysis. Our only concern is that the
signal $\delta T/T$ (whatever it consists of) is measured well compared to the
noise in a CMB map.} and the second term is the tSZ contribution, entering with a
different frequency dependence and parameterized by the Compton $y$-parameter,
\begin{align}
  y = \int n_\mathrm{e} \frac{k_\mathrm{B} \sigma_\mathrm{T} T_\mathrm{e}}{m_\mathrm{e} c^2}\ \mathrm{d}\mathrm{s}.
  \label{eq:comptony}
\end{align}
Here $m_\mathrm{e}$ the electron mass, $\sigma_\mathrm{T}$ the
Thomson cross-section, $\mathrm{ds}$ the differential distance along the line of
sight $\vec{\hat{n}}$, and $n_\mathrm{e}$ and $T_\mathrm{e}$ are the electron number density and
temperature. Next we apply a Lorentz boost ($\bm{\beta}\equiv \vec{v}/c$) from the unprimed CMB frame into the primed observation or solar-system frame to obtain
\begin{align}
  \delta I'(\vec{\hat{n}}') &= I'f(x')\left[\frac{\delta T'(\vec{\hat{n}}')}{T_0'} +
  y'(\vec{\hat{n}}')Y(x')\right].
  \label{eq:boosted}
\end{align}
Here $T_0'$ is the new boosted blackbody temperature, and only differs from $T_0$ to lowest order by $\beta^2$,
\begin{align}
  T_0'=T_0+\frac{\beta^2}{2}T_0;
\end{align}
thus to first order $T_0'=T_0$.
Taking each piece in turn, this transforms as (to first order)
\begin{align}
  I'f(x') &= If(x)\left[1 + \beta\mu (Y(x) + 3\beta\mu)\right], \\
  \frac{\delta T'(\vec{\hat{n}}')}{T_0'} &= \frac{\delta
  T(\vec{\hat{n}}')}{T_0} + \beta\mu, \\
  y'(\vec{\hat{n}}')Y(x') &= y(\vec{\hat{n}}')\left[Y(x) - \beta\mu x \frac{dY(x)}{dx}\right], \\
  \vec{\hat{n}}' &= \vec{\hat{n}} - \nabla(\vec{\hat{n}}\cdot\bm{\beta}),
  \label{eq:pieces}
\end{align}
where $\mu=\cos\theta$, and $\theta$ is defined as the angle from the direction
$\bm{\beta}$ to the line of sight.

\begin{figure*}[htbp!]
\mbox{\includegraphics[width=0.5\hsize]{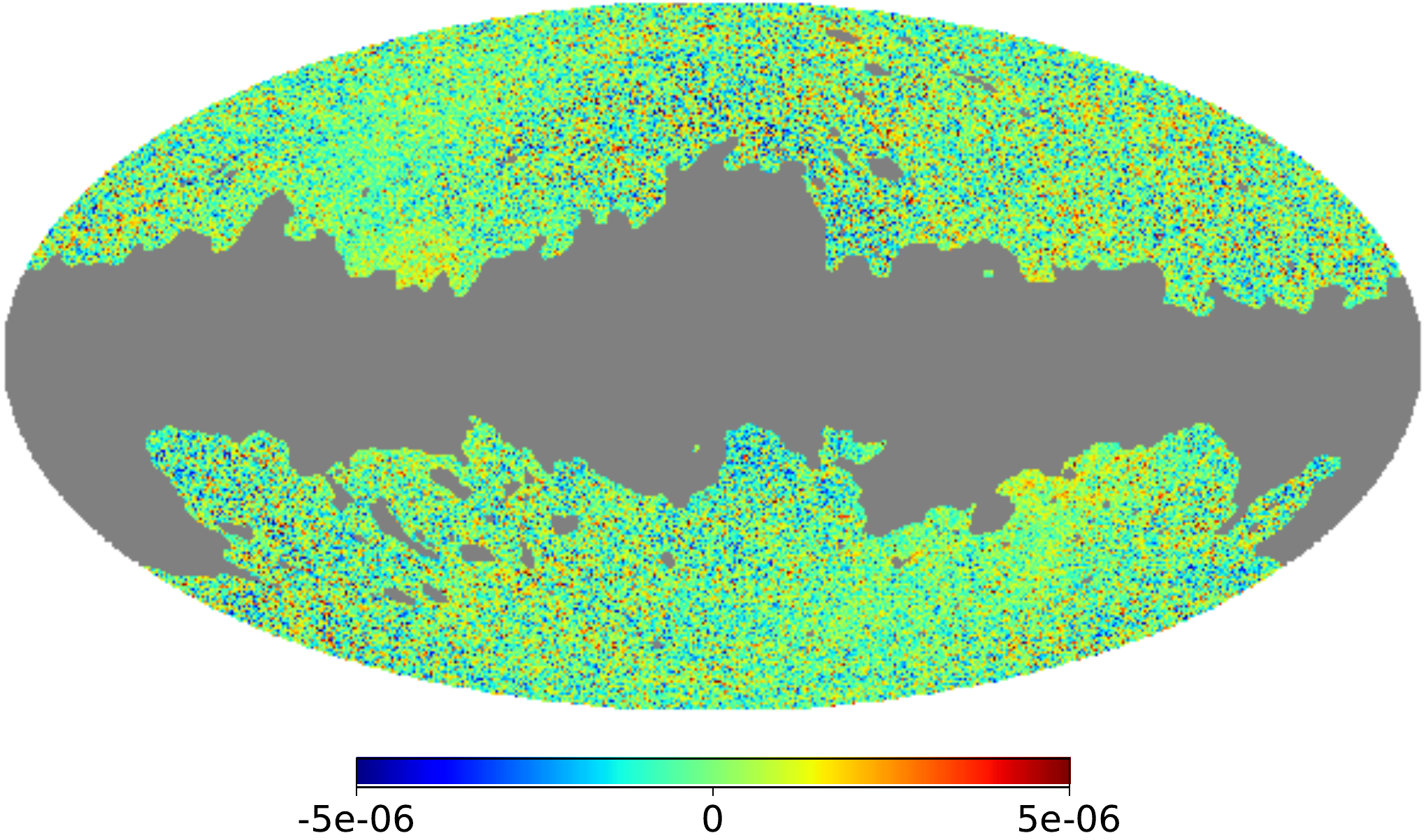}}
\mbox{\includegraphics[width=0.5\hsize]{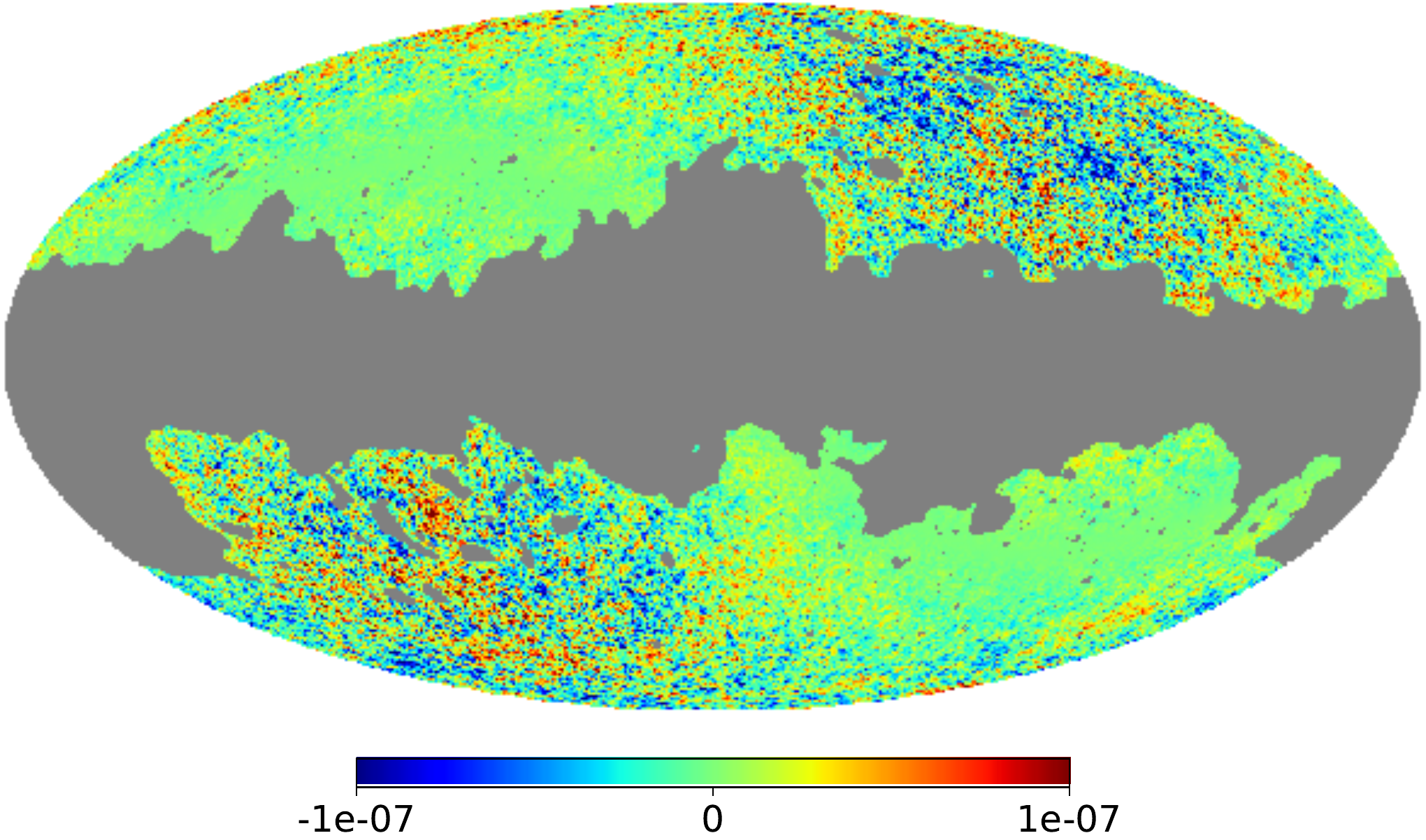}}
\mbox{\includegraphics[width=0.5\hsize]{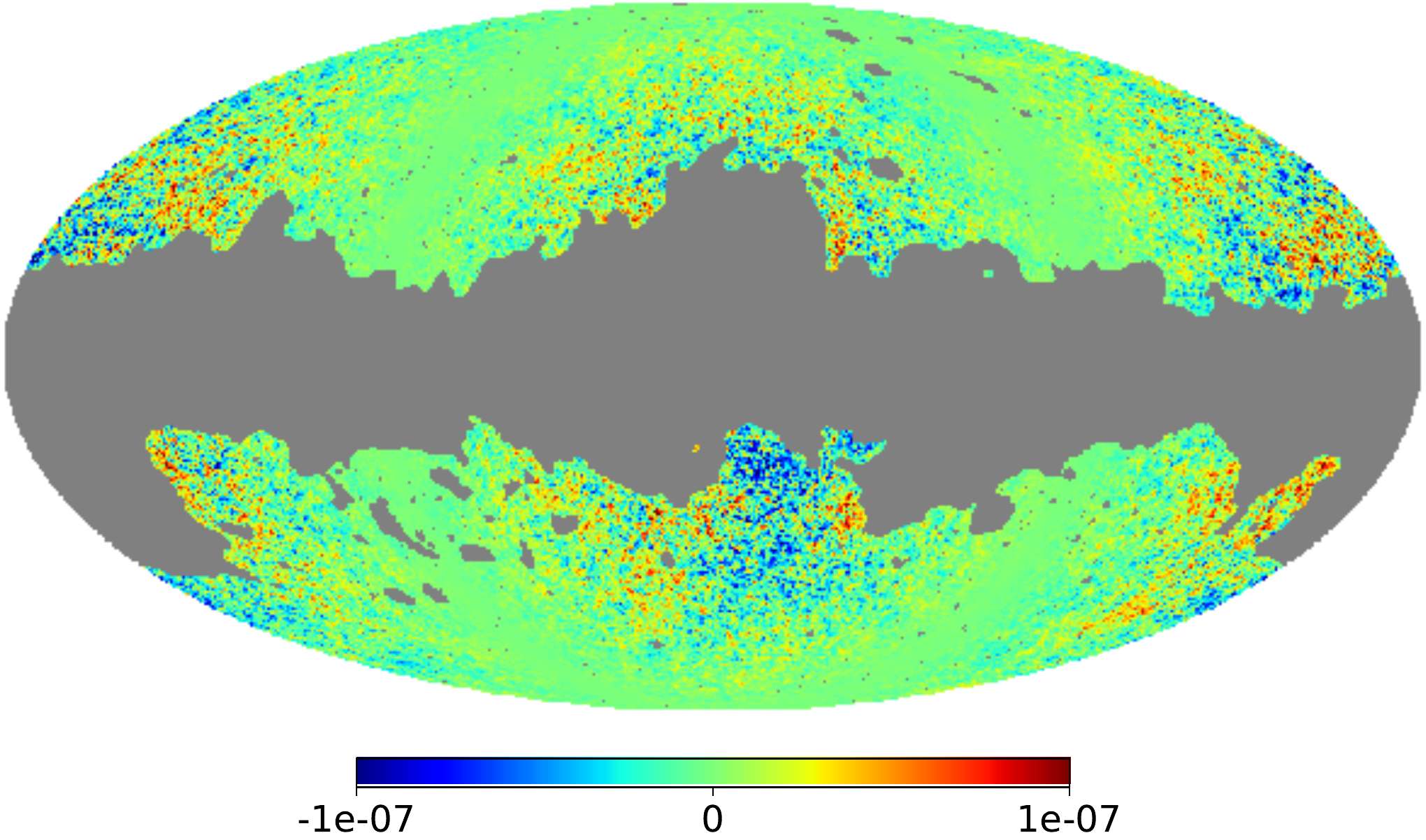}}
\mbox{\includegraphics[width=0.5\hsize]{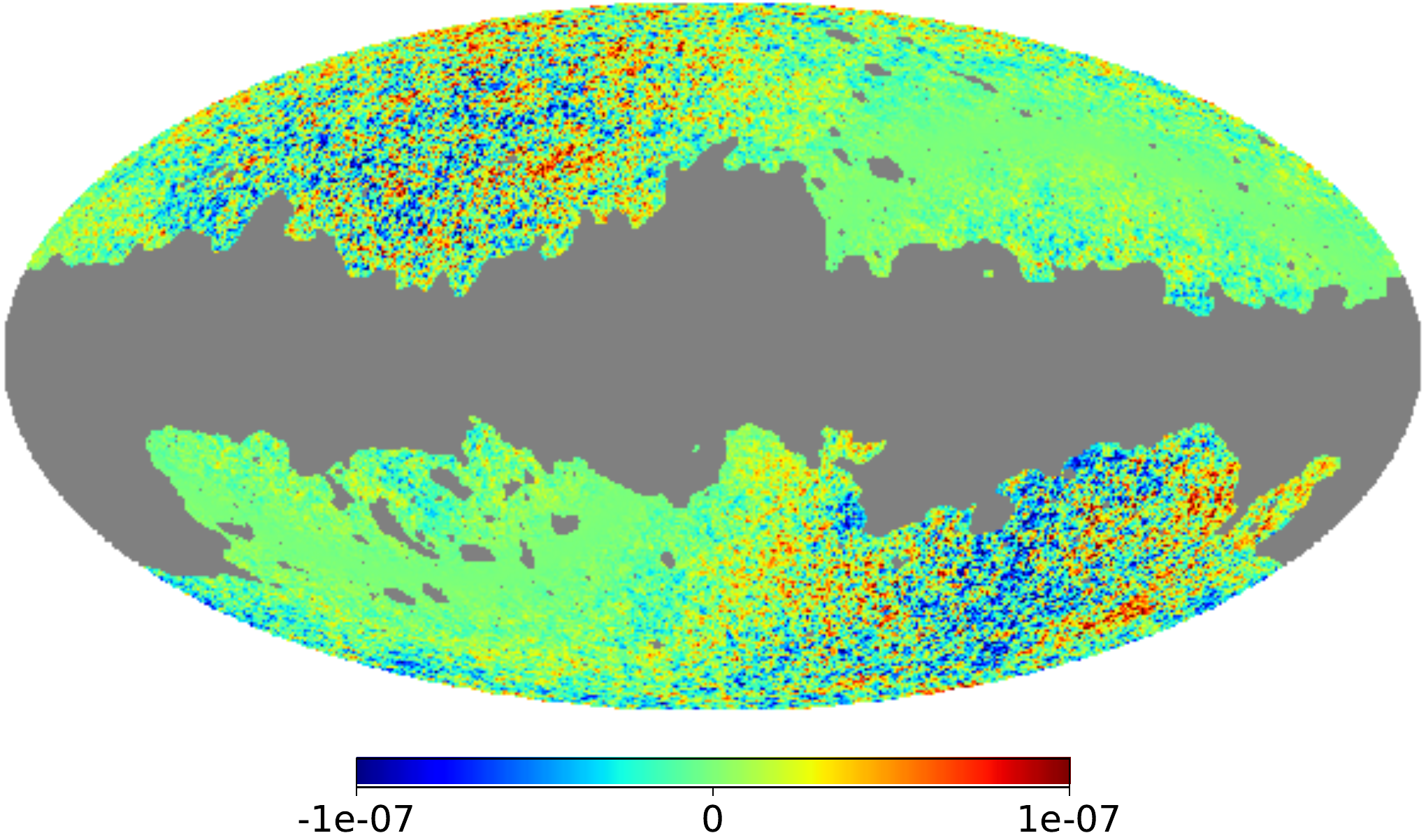}}
\caption{Map of the tSZ effect from the \milca\ component-separation method in $y$-map units (top left) and the expected
modulated CMB signal (top right) generated using the \smicano\ CMB map in units of $T_0$. The bottom left and right figures are the CMB anisotropies modulated in orthogonal directions to the CMB dipole, as explained
in the text after Eq.~\eqref{eq:template}. The greyed out region shows the mask
used for our analysis. \nilc{}\ and \twodilc{}\ $y$~maps, as well as \twodilc{}\ CMB modulated anisotropies, are not shown, since they look essentially the same as the maps presented here. Note that the map of the tSZ effect (top left) has a different scale bar when compared to the other three (i.e.,\ the modulation
signal is about 50 times weaker).}
\label{fig:template_data}
\end{figure*}

Equation~\eqref{eq:boosted} can thus be written as \footnote{We have left the primes on the $\hat{n}'$s as a matter of convenience; expanding this further would explicitly show the aberration effect, which we do not explore in this analysis.}

\begin{align}
    \delta I'(\vec{\hat{n}}')&=I f(x)\Bigl(1 + \beta\mu Y(x) + 3\beta\mu\Bigr)\times
  \nonumber\\
  &\Bigl( \frac{\delta T(\vec{\hat{n}}')}{T_0}
  + \beta\mu+y(\hat{\vec{n}}')Y(x) - y(\hat{\vec{n}}')\beta\mu x
  \frac{dY(x)}{dx}\Bigr)
    \label{eq:anisosimple}
\end{align}
or more explicitly, to first order in $\beta$,
\begin{align}
  \frac{\delta I'(\vec{\hat{n}}')}{If(x)} =\,& \frac{\delta T(\vec{\hat{n}}')}{T_0} + \beta\mu \left[1 + 3 \frac{\delta T(\vec{\hat{n}}')}{T_0} \right] \notag \\
  &+ Y(x) \left[y(\vec{\hat{n}}') + \beta\mu \frac{\delta T(\vec{\hat{n}}')}{T_0} \right] \notag \\
  &+ \beta\mu y(\vec{\hat{n}}') \left[3Y(x) + Y^2(x) - x \frac{dY(x)}{dx}\right],
  \label{eq:ianisotropiesexplicit}
\end{align}
where we have split up each line on the right-hand side according to the
frequency dependence. Assuming perfect component separation, and comparing with
Eq.~\eqref{eq:firstorder}, the first line
of Eq.~\eqref{eq:ianisotropiesexplicit}
shows that the boost induces a pure dipole ($\beta\mu$), an aberration effect
($\delta T(\vec{\hat{n}}')/T_0 - \delta T(\vec{\hat{n}})/T_0$), and a dipolar
modulation ($3\beta\mu \delta T/T_0$) of the CMB. The first effect is the classical CMB dipole, which has been
measured many times most
recently by \Planck\ \citep{planck2016-l01},
with the highest accuracy so far achieved. The effects of aberration and dipolar modulation
(both frequency-independent and frequency-dependent parts) were measured in
\citet{planck2013-pipaberration} at a combined significance level of
$5\,\sigma$.

In the second line of Eq.~\eqref{eq:ianisotropiesexplicit} we see that the boost also induces a change in
a map of the tSZ effect. The original $y$ signal is aberrated
($y(\vec{\hat{n}}') - y(\vec{\hat{n}})$) and also gains a contribution from the dipolar
modulated CMB ($\beta\mu\delta T/T_0$). This last effect is what we measure in
this paper for the first time.  Its expected signal can be seen in
\fig\ref{fig:template_data} (top right panel), along with the full $y$~map obtained via the
\milca\ method \citep[][\fig\ref{fig:template_data} top left]{Hurier2013}. It is worth
noting that although the contribution to the tSZ map is a dipolar modulation of
the CMB \emph{anisotropies}, this induces power in the $y$~map that is modulated like a
\emph{quadrupolar} pattern (due to the lack of correlation between the CMB
anisotropies and $y$ signal). That is, a $y$~map contains more power in the
poles of the dipole, relative to the corresponding equator (see
\fig\ref{fig:template_data}). It should be possible to pull out this signal compared with modulation patterns oriented in orthogonal directions (lower panels of \fig\ref{fig:template_data}).

We note that the final line in Eq.~\eqref{eq:ianisotropiesexplicit} is simply
the dipole modulation of the tSZ effect, with a peculiar frequency dependence
\citep[][]{Chluba2005}. In principle one could generate a map of the
anisotropies in this new frequency dependence and use the known CMB dipole to measure the
$y$ anisotropies again. Such a measurement would be correlated with the original
$y$~map, but would have independent noise properties and also have a very low amplitude. From a practical
perspective it is unlikely that such a measurement would yield any significant
information increase, since the signal is contaminated with relativistic tSZ and
kSZ effects, and is suppressed by a factor of $\beta$ \citep[][]{Chluba2005}.

\section{Data}
\label{sec:data}
We look for the signal by cross-correlating a template map derived
from the CMB temperature data with a $y$~map. Therefore it is important that
the CMB map is free of $y$ residuals and that the $y$~map is free of CMB
residuals, in order to avoid spurious correlations.

To this end, we use the so-called \twodilc\ CMB temperature map \citep[first
used for kSZ detection\footnote{The map name "\twodilc"\ was adopted because of
the two-dimensional (2D) constraint imposed on the internal linear combination
(ILC) weights of being aligned with the CMB/kSZ spectrum while being orthogonal
to the tSZ spectrum.} in][]{planck2013-XIII}, which was produced by the
``Constrained ILC'' component-separation method designed by
\cite{Remazeilles2011} to explicitly null out the contribution from the
$y$-type spectral distortions in the CMB map. We also use the \smicano\ temperature map, similarly produced with express intent of removing the $y$-type spectral distortions, and which was generated with the \Planck\ 2018 data release \citep{planck2016-l04}.
Likewise, we use the corresponding \twodilc\ $y$~map, and the \Planck\ \milca\ $y$~map, which explicitly null out the contributions from a (differential) blackbody spectral distribution in the $y$~map \citep{Hurier2013, planck2014-a28}. We also consider the \Planck\ \nilc\ $y$~map,
 which does not explicitly null out the blackbody contribution.
The extra constraint to remove CMB anisotropies in the \twodilc\ $y$~map, or in the \milca\ $y$~map, is at the expense of leaving more
contamination by diffuse foregrounds and noise. In
\cite{planck2014-a28}, the \Planck\ \nilc\ $y$~map was preferred over the
\twodilc\ $y$~map to measure the angular power spectrum of tSZ anisotropies,
the CMB contaminant being negligible compared to diffuse foregrounds.
Conversely, to measure the dipole modulation of the CMB anisotropies
in the $y$~map, the \twodilc\ and \milca\ $y$~maps are preferred over the \Planck\ \nilc\
$y$~map because the latter does not fully null out the
contribution from the CMB. This significantly contaminates the signal we are looking for, as can be seen in Appendix~\ref{app:nilc}. It is noted in \cite{planck2014-a28} that for $\ell>2000$ the signal is dominated by correlated noise, and so we use the same cut as used in their analysis of $\ell_{max}=1411$, this is further justified in Sect.~\ref{sec:pscl}.

Figure~\ref{fig:template_data} shows the mask used in our analysis. This is the union of the \Planck\ 2018 data release common temperature confidence mask \citep[][]{planck2016-l04}, and the corresponding $y$-map foreground masks \citep[][]{planck2014-a28}. This was then extended by $1\deg$ and apodized with a $200\arcm$ Gaussian beam. To account for any masked sections lost during the smoothing, the original mask was then added back. Tests were also done using the $y$-map point-source mask, with negligible changes seen in the results, and was thus omitted from the final analysis. This procedure aims to allow for the maximum signal while minimizing the foreground contamination. Various combinations of mask sizes and apodizations were also tested and final results were consistent, independent of the choice of mask.

\section{Analysis}
\label{sec:analysis}
From Eq.~\eqref{eq:ianisotropiesexplicit} we see that a map of the tSZ effect ($\mathcal{M_{\mathrm{SZ}}}$)
contains the following terms (for each pixel, or direction $\vec{\hat{n}}$):
\begin{align}
  \mathcal{M_{\mathrm{SZ}}}=y + \eta^y + \beta\mu \frac{\delta T}{T_0},
  \label{eq:ymap}
\end{align}
where $\eta^y$ is simply the noise in the $y$~map, and we have neglected the
aberration effect. Our goal is to isolate the final term in
Eq.~\eqref{eq:ymap}, which we do via a suitable cross-correlation with a CMB
map. A map of the CMB ($\mathcal{M}_{\mathrm{CMB}}$) contains the following terms:
\begin{align}
  \mathcal{M_{\mathrm{CMB}}}=\frac{\delta T}{T_0} + \eta^T + 3\beta\mu \frac{\delta T}{T_0},
  \label{eq:cmbmap}
\end{align}
where we have explicitly removed the full dipole term, and $\eta^T$ is the noise
in the CMB map. If we multiply our CMB map (Eq.~\ref{eq:cmbmap}) with
$\beta\mu$ and cross-correlate that with our tSZ map (Eq.~\ref{eq:ymap}), then
we can directly probe the dipole modulation. This
of course neglects the noise and modulation terms in the CMB map, which we are
justified in doing because the noise term is sub-dominant, except at very small
scales (we make the restriction $\ell_{\max} = 1411$, so that the $y$~map and CMB maps are
still signal dominated \citealt{planck2014-a28}), and because the modulation term becomes second order in
$\beta$. Equivalently one could directly cross-correlate
Eq.~\eqref{eq:ymap} with Eq.~\eqref{eq:cmbmap} and look for the signal in harmonic space from the
coupling of $\ell$ and $\ell\pm 1$ modes.

 In \cite{planck2013-pipaberration} a quadratic estimator was
used to determine the dipole aberration and modulation, in essence using the auto-correlation of the CMB fluctuation
temperature maps, weighted appropriately to extract the dipole signal. The auto-correlation naturally introduces a
correlated noise term, which must be well understood for this method to work.
In this paper we take advantage of the fact that we know the true CMB
fluctuations with excellent precision and therefore the signal that should be
present in the $y$ map. We can therefore exploit the full angular dependence of
the modulation signal and remove much of the cosmic variance that would be
present in the auto-correlation.

In order to implement this idea we define three templates, $B_i$ (with
$i=1,2,3$) as
\begin{align}
  B_i(\hat{\vec{n}}) &= \beta \hat{\vec{n}} \cdot \hat{\vec{m}}_i \frac{\delta T}{T_0} (\hat{\vec{n}}),
  \label{eq:template}
\end{align}
where $\beta = v/c$ is $1.23357\times10^{-3}$ \citep{planck2016-l01} and $\hat{\vec{m}}_1, \hat{\vec{m}}_2,
\hat{\vec{m}}_3$ are the CMB dipole direction, an orthogonal direction in the
Galactic plane, and the third remaining orthogonal direction \citep[see
\fig\ref{fig:template_data} and][for a similar
approach]{planck2013-pipaberration}. Note that in Eq.~\eqref{eq:template}, we
simply use our CMB map in place of $\delta T/T_0$.
We use two distinct methods to accomplish this, discussed in
detail in Sects.~\ref{sec:stacking} and \ref{sec:pscl}.

In the region where the CMB is signal dominated we can regard $\delta T/T_0$ as
fixed, and thus our templates $B_i$ are fixed. Due to the presence of the CMB
dipole, the signal $B_1$ should be present in the $y$~map. We can therefore
directly cross-correlate $B_1$ with our $y$~map (Eq.~\ref{eq:ymap}) to pull out
the signal. Likewise, the cross-correlation of $B_2$ and $B_3$ with our $y$~map
should give results consistent with noise, although the coupling with the noise and mask leads to a bias that is recovered through simulations.

Our $y$ simulations are generated by first computing the power spectra of our
data $y$~maps; specifically we apply the {\tt MASTER} method using the {\tt NaMASTER} routine \citep{NaMaster} to account for the applied mask \citep[][]{Hivon2002}. Then we
generate $y$~maps using this power-spectrum with the \healpix\
\citep[][]{gorski2005} routine {\tt synfast}.\footnote{Note this means
that our simulations contain no non-Gaussianities, unlike the real SZ data; however, this should have no effect on the power spectrum, since non-Gaussianities are only detectable at higher order such as the bispectrum \citep[see e.g.,][]{lacasa2012characterization}. For further discussion of see Appendix~\ref{sec:bispectrum}.}
This is done separately for the \twodilc\ and \milca\ maps because they have
different noise properties (and thus different total power spectra).
For our simulations that include the dipolar modulated CMB anisotropies we add
the last term of Eq.~\eqref{eq:ymap}. We finally apply a Gaussian smoothing of
5\arcm\ to model the telescope beam.

For each analysis method (to be described in the following subsections) we
estimate the amplitude of the dipole ($\hat{\beta}_i$) in each of the three
orthogonal directions.\footnote{Note that $\hat{\beta}_i$ is used here to denote the estimator, not a unit vector.} We apply the same analysis on a suite of 1000 $y$
simulations, generated with and without the dipolar modulation term in
Eq.~\eqref{eq:ymap}. We are then able to generate a covariance that
appropriately contains the effects of the mask we use and are able to compute
any bias that the mask induces. On the assumption (verified by our simulations)
that the $\hat{\beta}_i$ estimators are Gaussian, we are able to compute a
value of $\chi^2$ for the case of no CMB term and with the CMB term (see
Table~\ref{tab:chi2}). We can then apply Bayes' theorem along with our
covariance to calculate the probability that each model is true (with or
without the CMB term) and the posterior of our dipole parameters ($\beta, l,
b$), summarized in Table~\ref{tab:chi2} and \fig\ref{fig:triangle}.

We estimate the covariance $C_{ij}$ of the $\hat{\beta}_i$ using the
simulations\footnote{It makes no appreciable difference whether we use the
simulations with or without the dipole term to calculate the covariance.} and
calculate the $\chi^2$ as
\begin{align}
  \chi^2_{k} &= \sum_{ij} \left(\hat{\beta}_i - \left<\hat{\beta}_i\right>_k\right) C^{-1}_{ij}
  \left(\hat{\beta}_j - \left<\hat{\beta}_j\right>_k\right),
  \label{eq:chi2}
\end{align}
where $k$ denotes whether the expectation value in the sum is taken over the
simulations that do or do not include the CMB term. For definiteness we define
the null hypothesis $H_0$ ($k=0$) to not include the CMB term, while hypothesis
$H_1$ ($k=1$) does include the CMB term. We can then directly calculate the
probability that $H_k$ is true given the data ($\hat{\beta}_i$) as
\begin{align}
  P(H_k|\hat{\beta}_i) &= P(\hat{\beta}_i|H_k) P(H_k),
  \label{eq:probofhypo} \\
  P(\hat{\beta}_i|H_k) &= \frac{1}{\sqrt{|2\pi C|}} e^{-\chi^2_k/2}.
  \label{eq:like}
\end{align}
We can calculate the odds ratio, $O_{10}$, on the assumption that the two
hypotheses are equally likely,
\begin{align}
  O_{10} &\equiv \frac{P(H_1|\hat{\beta}_i)}{P(H_0|\hat{\beta}_i)} =
  \frac{e^{-\chi^2_1/2}}{e^{-\chi^2_0/2}}.
  \label{eq:oddsratio}
\end{align}
This quantity tells us to what degree $H_1$ should be trusted over $H_0$.
Assuming that the two hypotheses are exhaustive, it is directly related to the
probability that the individual hypotheses are true:
\begin{align}
  P(H_0|\hat{\beta}_i) &= \frac{1}{1 + O_{10}}, \\
  P(H_1|\hat{\beta}_i) &= \frac{O_{10}}{1 + O_{10}}.
  \label{eq:prob_oddsratio}
\end{align}
These quantities and the $\chi^2_k$ values are given in Table~\ref{tab:chi2}.

We can also generate a likelihood for our parameters with the same covariance
matrix:
\begin{align}
  \mathcal{L}(\beta_i) &= \frac{1}{\sqrt{|2\pi C|}} e^{-\chi^2/2},
  \label{eq:likelihood}
\end{align}
where we define the modified $\chi^2$ as above.
We can then apply Bayes' theorem with uniform priors on the $\beta_i$, equating
the posterior of $\beta_i$ with Eq.~\eqref{eq:likelihood}. A simple conversion
allows us to obtain the posterior of the parameters in spherical coordinates
$(\tilde{\beta}, \tilde{l}, \tilde{b})$. We show this in
\fig\ref{fig:triangle} for our two analyses, using the \twodilc\ and \milca\
maps.

In the following subsections we describe two methods of cross-correlation: the
first we perform directly in map-space; and the second is performed in harmonic
space. An advantage of using two independent methods is that their noise properties are different; for example, working in harmonic space introduces complications with masking, whereas in map space, although it may not be clear how to optimally weight the data, the estimator has less sensitivity to large-scale systematic effects.
Thus, the advantage of using two approaches will become apparent when we try to assess the level of systematic error in our analysis.

\subsection{Map-space method}
\label{sec:stacking}
First we apply our mask to the templates $B_i$ and $y$~map. Then we locate all
peaks (i.e., local maxima or minima) of the template map $B_i$ and select a patch of radius $2\pdeg0$ around
each peak. Our specific implementation of the peak method follows earlier studies, for example \citet{planck2016-l07}.  The weighting scheme has not been shown to be optimal, but a similar approach was used for determining constraints on cosmic birefringence \citet{Birefringence} and gave similar results to using the power spectra (and the issue of weighting is further discussed in \citealt{2019Saddle}). Intuitively we would expect that sharper peaks have a higher signal, and hence that influences our choice for the weighting scheme described below. For every peak we obtain an estimate of $\hat{\beta}_i$ by the simple
operation
\begin{align}
  \hat{\beta}_{i,p} &= \beta \frac{\sum_{k\in D(p)} B_{i,k} y_k}{\sum_{k\in D(p)}
  B_{i,k}^2},
  \label{eq:mapcross}
\end{align}
where $D(p)$ is the collection of all \emph{unmasked} pixels in a $2\pdeg0$
radius centred on pixel $p$, and $p$ is the position of a peak.
Equation~\eqref{eq:mapcross} is simply a cross-correlation in map space and by itself
offers a highly-noisy (and largely unbiased\footnote{This is strictly true
for $\hat{\beta}_1$ only; 
the presence of a strong signal in the data is
correlated in orthogonal directions due to the mask and thus may appear as a
mild bias in $\hat{\beta}_2$ and $\hat{\beta}_3$. There is also a bias due to the correlations between the templates. The weighting in harmonic space is much simpler, and so this effect is taken into account in the harmonic-space method; however, due to the complicated nature of the weighting in the map-space method it is not included in this section. We discuss this further in
Sect.~\ref{sec:results}.}) estimate.

We then combine all
individual peak estimates with a set of weights ($w_p$) to give our full
estimate:
\begin{align}
  \hat{\beta}_i &= \frac{\sum_p w_{i,p} \hat{\beta}_{i,p}}{\sum_p w_{i,p}}.
  \label{eq:mapweighting}
\end{align}

The values of $w_{i,p}$ depend solely on the templates $B_i$, and they can be
chosen to obtain the smallest uncertainties. We
choose $w_p$ to be proportional to the square of the dipole, which ensures that
peaks near the dipole direction (and anti-direction) are weighted more than
those close to the corresponding equator. We further choose that the weights
are proportional to the square of the Laplacian at the peak
\citep{Desjacques2008}; this favours sharply defined peaks over shallow ones.
Finally we account for the scan strategy of the \Planck\ mission by weighting
by the 217-GHz hits map \citep[][denoted $H^{217}_p$]{planck2014-a09}, though this choice provides no appreciable difference to our results. The weights then are explicitly
\begin{align}
  w_{i,p} &= |\hat{\vec{n}} \cdot \hat{\vec{m}}_i|^2_p \left(\left.\nabla^2(B_i)\right|_p\right)^2 H^{217}_p.
  \label{eq:mweights}
\end{align}
We evaluate the Laplacian numerically in pixel space at pixel $p$. The
weighting scheme closely resembles the bias factors that come about when
relating peaks to temperature fluctuations \citep{Bond1987}, as used in
\citet{komatsu2010}, \citet{planck2014-a23} and \citet{Jow2019}.

Combining Eqs.~\eqref{eq:mapweighting} and~\eqref{eq:mweights} gives us our
estimates, $\hat{\beta}_i$. We apply the method for each of our simulated
$y$~maps, in exactly the same way as for the data.

\subsection{Harmonic-space method}
\label{sec:pscl}

The alternative approach is to directly cross-correlate Eq.~\eqref{eq:template}
with the $y$~map, and compare this to the auto-correlation of
Eq.~\eqref{eq:template}.

Our first step is identical to the previous method in that we mask the
templates $B_i$ and $y$~maps. Under the assumption that the $y$~map contains
the template ($B_i$), the $y$ multipoles are Gaussian random numbers with mean
and variance given by
\begin{align}
  s^i_{\ell m} &= \int d\Omega\, \beta\, \hat{\vec{m}}_i\cdot\hat{\vec{n}}\, \frac{\delta T}{T_0} M(\Omega) Y^*_{\ell m},\\
  \sigma^2_{\ell} &= C^y_{\ell} + N^y_{\ell},
  \label{eq:mean_signal}
\end{align}
respectively, where $M(\Omega)$ is the mask over the sphere, $Y_{\ell m}$ are the spherical harmonics, and the $\hat{\vec{m}}_i$ are as defined in Eq. \eqref{eq:template}. Thus we can obtain an estimate of
$\beta_i$ by taking the cross-correlation with inverse-variance weighting. We
can demonstrate this simply by writing our $y$~map as a sum of our expected
signal plus everything else\footnote{Note that here $\eta^y_{\ell m}$ is different to that in Eq.~\eqref{eq:ymap}, since it now also includes the $y$ signal, which is treated as a noise term in this analysis.},
\begin{align}
  y_{\ell m} &= \frac{\beta_i}{\beta} s^{i}_{\ell m} + \eta^y_{\ell m}.
  \label{eq:yderivation}
\end{align}
Here our signal is of course given when $\beta_i = \beta \delta_{1 i}$ and all
sources of noise, such as tSZ, are given by $\eta^y_{\ell m}$. We then
cross-correlate with our template and sum over all multipoles with
inverse-variance weighting.  We explicitly consider noise in our template,
that is our template ($\bar{s}^i_{\ell m}$) is related to
Eq.~\eqref{eq:mean_signal} via, $\bar{s}^i_{\ell m} = s^i_{\ell m} +
\eta^t_{\ell m}$, where $\eta^t_{\ell m}$ is the noise in our template. Then
the cross-correlation looks like,
\begin{align}
  \sum_{\ell m} \bar{s}^{i'}_{\ell m} y^*_{\ell m}/ \sigma^2_\ell &=
  \frac{\beta_i}{\beta} \sum_{\ell m} \bar{s}^{i'}_{\ell m} (s^i_{\ell m})^* /
  \sigma^2_\ell \notag\\
  &+ \sum_{\ell m} \bar{s}^{i'}_{\ell m} (\eta^y_{\ell m})^*
  /\sigma^2_\ell,
\end{align}
and expanding it out, this becomes
\begin{align}
  \sum_{\ell m} s^{i'}_{\ell m} y^*_{\ell m}/ \sigma^2_\ell + \sum_{\ell m}
  \eta^{t}_{\ell m} y^*_{\ell m}/ \sigma^2_\ell &=
  \frac{\beta_i}{\beta} \sum_{\ell m} s^{i'}_{\ell m} (s^i_{\ell m})^* /
  \sigma^2_\ell \notag\\
  &+ \frac{\beta_i}{\beta} \sum_{\ell m} \eta^{t}_{\ell m} (s^i_{\ell m})^* /
  \sigma^2_\ell \notag\\
  &+ \sum_{\ell m} s^{i'}_{\ell m} (\eta^y_{\ell m})^*
  /\sigma^2_\ell \notag\\
  &+ \sum_{\ell m} \eta^{t}_{\ell m} (\eta^y_{\ell m})^*
  /\sigma^2_\ell.
  \label{eq:hcrossderiv}
\end{align}
The last term on the left and last three terms on the right are all
statistically zero, since our template does not correlate with tSZ or any other
types of noise and the noise in our template does not correlate with the
template itself or with noise in the $y$~map (by assumption).  Hence we can
solve for $\beta_i$ neglecting those terms, to produce our estimator
$\hat{\beta}_i$:
\begin{align}
  \hat{\beta}_i &= \beta \sum_{i'} \left[\sum_{\ell m}^{\ell_{\max}}
  s^i_{\ell m}(s^{i'}_{\ell m})^* / \sigma^2_{\ell}\right]^{-1} \sum_{\ell
  m}^{\ell_{\max}} s^{i'}_{\ell m}
  (y_{\ell m})^*/\sigma^2_{\ell}.
  \label{eq:hcross}
\end{align}
It is important to note that in practice we do not have $s^i_{\ell m}$, since we do not know the
exact realization of noise in the CMB, so we instead use $\bar{s}^i_{\ell m}$. 
Using Weiner-filtered results would allow us to calculate $s^i_{\ell m}$, but adds complexity in the masking process.
We can compare what the Weiner-filtered results would be,
\begin{align}
	\sum_{\ell}\frac{(C_{\ell}^{TT})^2}{C_{\ell}^{TT}+N_{\ell}^{TT}}\frac{(2\ell+1)}{\sigma_\ell^2},
	\label{eq:weinerfil-result}
\end{align}
to our results,
\begin{align}
 	\sum_{\ell}(C_\ell^{TT}+N_\ell^{TT})\frac{(2\ell+1)}{\sigma_\ell^2}
 	\label{eq:hsmresults}
\end{align}
and find the bias to be on the order of 2\,\% for $\ell_{\max} = 1411$, justifying our use of the cut-off.

Equation~\eqref{eq:hcross} is in fact a direct solution for $\beta_i$ in the absence
of noise, since it is the direct solution of Eq.~\eqref{eq:hcrossderiv} in the
absence of noise.

Relating back to the map-space method, $s^i_{\ell m}$ are the spherical harmonic coefficients of the templates denoted previously by $B_i$, and $y_{\ell m}$ are the spherical harmonic coefficients of the $y$~map. The values for $s^i_{\ell m}(s^{i'}_{\ell,m})^*$ and $s^i_{\ell m} (y_{\ell m})^*$ may be computed using the maps with the \healpix\ \citep[][]{gorski2005} routine {\tt anafast}.
In the case of the term $s^i_{\ell m}(s^{i'}_{\ell,m})^*$ this results in a $3\times3$ matrix for each $\ell$, with the cross-power spectrum for the three templates on the off diagonals.

In the
absence of a mask $M$ the signal $|s_{\ell m}|^2$ induces power in a
$\cos^2{\theta}$ pattern. The presence of a mask (being largely quadrupolar in
shape) induces power in a more complicated way, but has strong overlap with a
$\cos^2{\theta}$ pattern as well. Therefore the application of a mask
necessarily makes this method sub-optimal; however, since the template is
masked in the same way, the method is unbiased.

We apply the method for each of our simulated $y$~maps, in exactly the same way
as for the data, in order to assess whether the dipole modulation is
detected.

\section{Results}
\label{sec:results}

\begin{table}[htbp!]
\begingroup
\newdimen\tblskip \tblskip=5pt
\caption{Values of $\chi^2$ (with $N_{\rm dof} = 3$) under the assumption of no dipolar modulation term (``No dipole''), and assuming the presence of the dipolar modulation term (``With dipole'') for the \twodilc\ CMB template map. We include the probability that hypotheses of ``No dipole'' and ``With dipole'' are true. All data and analysis combinations are consistent with the dipole modulation term. The deviations range from 6.2 to $6.6\,\sigma$ for the harmonic-space analysis, and from 5.0 to $5.9\,\sigma$ for the map-space analysis.}
\label{tab:chi2}
\nointerlineskip
\vskip -3mm
\footnotesize
\setbox\tablebox=\vbox{
   \newdimen\digitwidth
   \setbox0=\hbox{\rm 0}
   \digitwidth=\wd0
   \catcode`*=\active
   \def*{\kern\digitwidth}
   \newdimen\signwidth
   \setbox0=\hbox{+}
   \signwidth=\wd0
   \catcode`!=\active
   \def!{\kern\signwidth}
\halign{\tabskip 0pt \hbox to 0.7in{#\leaderfil}\tabskip 6pt&
         \hfil#\hfil\tabskip 18pt&
         \hfil#\hfil\tabskip 18pt&
         \hfil#\hfil\tabskip 18pt&
         \hfil#\hfil\tabskip 0pt\cr
\noalign{\doubleline\vskip 3pt}
\omit& \multispan2\hfil No dipole\hfil& \multispan2\hfil With dipole\hfil\cr
\noalign{\vskip -3pt}
\omit& \multispan2\hrulefill& \multispan2\hrulefill\cr
\noalign{\vskip 4pt}
\omit\hfil Method\hfil& $*\chi^2$& $P(H_0|\hat{\beta}_i)$& $*\chi^2$& $P(H_1|\hat{\beta}_i)$\cr  
\noalign{\vskip 4pt\hrule\vskip 6pt}
\multispan5\hfil Harmonic-space analysis\hfil\cr
\noalign{\vskip 3pt}
\twodilc&  39.5& $4.0\times10^{-9*}$& 0.8& $1-4.0\times10^{-9*}$\cr
\noalign{\vskip 3pt}
\milca&    42.4& $8.4\times10^{-10}$& 0.7& $1-8.4\times10^{-10}$\cr
\noalign{\vskip 5pt\hrule\vskip 6pt}
\multispan5\hfil Map-space analysis \hfil\cr
\noalign{\vskip 3pt}
\twodilc&  38.6& $1.8\times10^{-8*}$& 3.0& $1-1.8\times10^{-8*}$\cr
\noalign{\vskip 3pt}
\milca&    24.8& $5.0\times10^{ -6}$& 0.4& $1-5.0\times10^{-6*}$\cr
\noalign{\vskip 5pt\hrule\vskip 4pt}}}
\endPlancktable                    
\endgroup
\end{table}

\begin{table}[htbp!]
\begingroup
\newdimen\tblskip \tblskip=5pt
\caption{As in Table~\ref{tab:chi2} but using \smicano\ CMB template maps.}
\label{tab:chi22}
\nointerlineskip
\vskip -3mm
\footnotesize
\setbox\tablebox=\vbox{
   \newdimen\digitwidth
   \setbox0=\hbox{\rm 0}
   \digitwidth=\wd0
   \catcode`*=\active
   \def*{\kern\digitwidth}
   \newdimen\signwidth
   \setbox0=\hbox{+}
   \signwidth=\wd0
   \catcode`!=\active
   \def!{\kern\signwidth}
\halign{\tabskip 0pt \hbox to 0.7in{#\leaderfil}\tabskip 6pt&
         \hfil#\hfil\tabskip 18pt&
         \hfil#\hfil\tabskip 18pt&
         \hfil#\hfil\tabskip 18pt&
         \hfil#\hfil\tabskip 0pt\cr
\noalign{\doubleline\vskip 3pt}
\omit& \multispan2\hfil No dipole\hfil& \multispan2\hfil With dipole\hfil\cr
\noalign{\vskip -3pt}
\omit& \multispan2\hrulefill& \multispan2\hrulefill\cr
\noalign{\vskip 4pt}
\omit\hfil Method\hfil& $*\chi^2$ & $P(H_0|\hat{\beta}_i)$& $*\chi^2$& $P(H_1|\hat{\beta}_i)$\cr   
\noalign{\vskip 5pt\hrule\vskip 6pt}
\multispan5\hfil Harmonic-space analysis\hfil\cr
\noalign{\vskip 3pt}
\twodilc&  41.9& $1.5\times10^{-9*}$& 1.2& $1-1.5\times10^{-9*}$\cr
\noalign{\vskip 3pt}
\milca&    45.4& $3.1\times10^{-10}$& 1.6& $1-3.1\times10^{-10}$\cr
\noalign{\vskip 5pt\hrule\vskip 6pt}
\multispan5 \hfil Map-space analysis \hfil\cr
\noalign{\vskip 3pt}
\twodilc&  40.1& $8.9\times10^{-9*}$& 3.0& $1-8.9\times10^{-9*}$\cr
\noalign{\vskip 3pt}
\milca&    27.9& $1.1\times10^{-6*}$& 0.4& $1-1.1\times10^{-6*}$\cr
\noalign{\vskip 5pt\hrule\vskip 4pt}}}
\endPlancktable                    
\endgroup
\end{table}

\begin{figure*}[htbp!]
\mbox{\includegraphics[width=0.5\hsize]{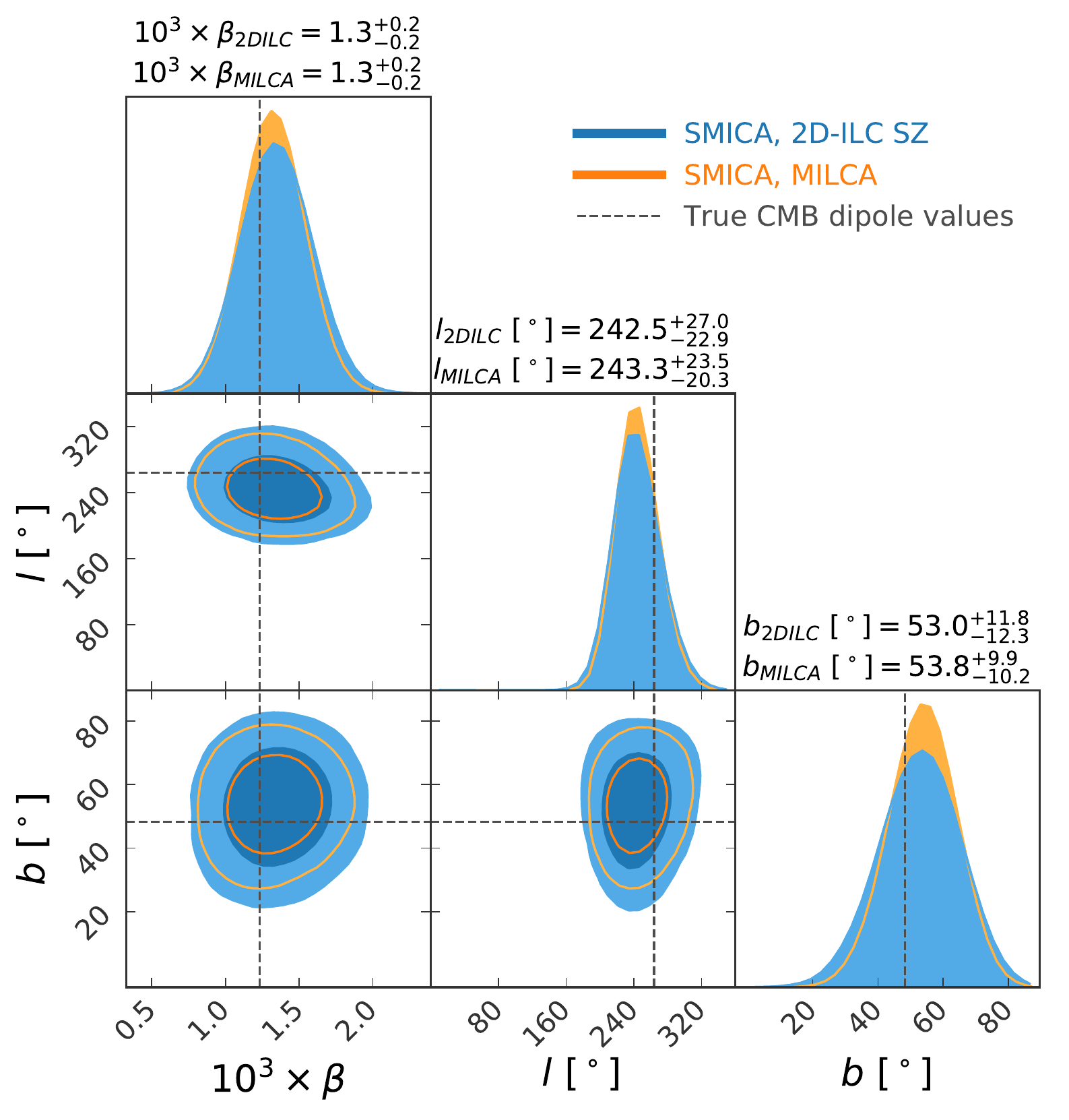}}
\mbox{\includegraphics[width=0.5\hsize]{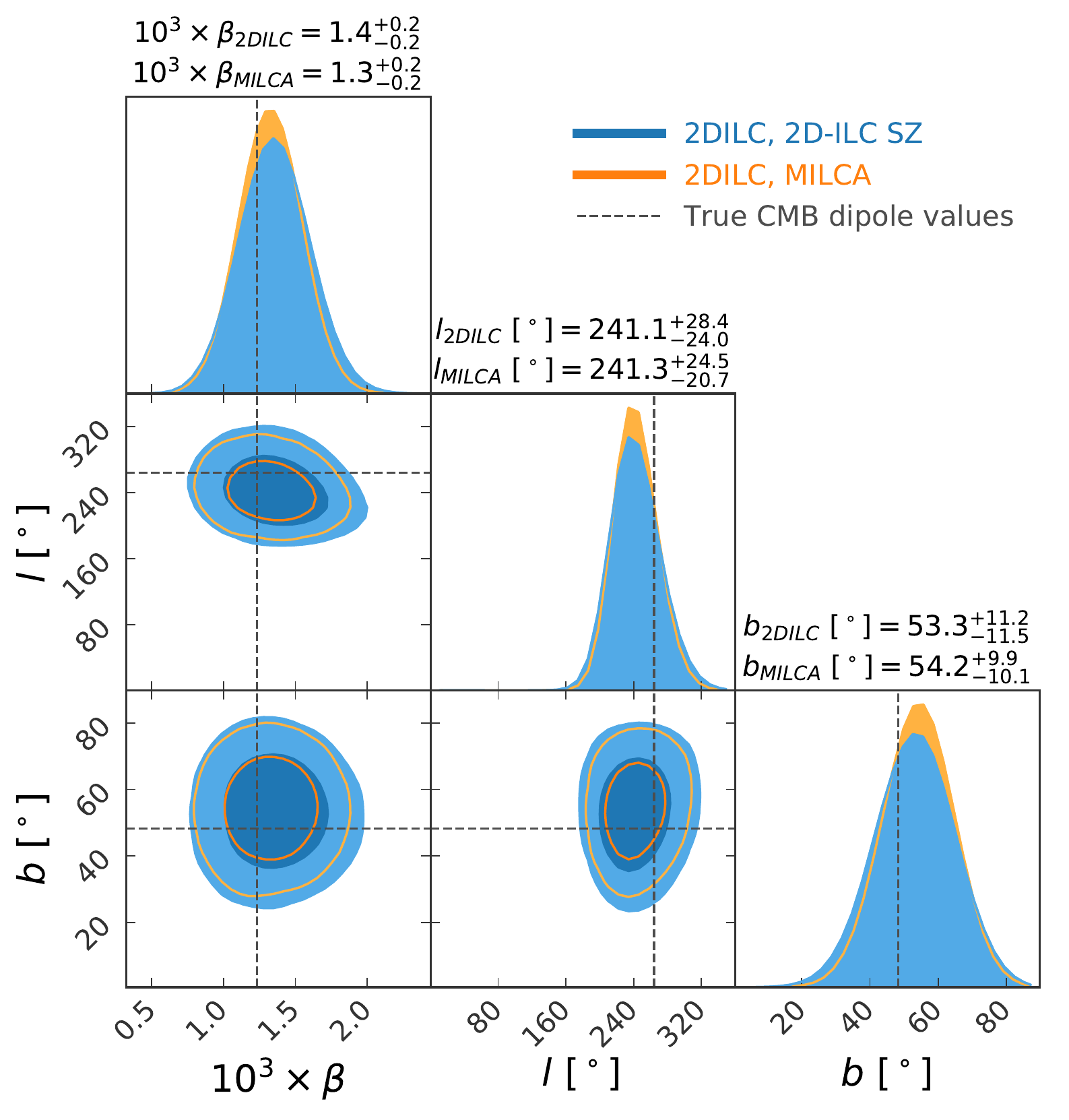}}
\mbox{\includegraphics[width=0.5\hsize]{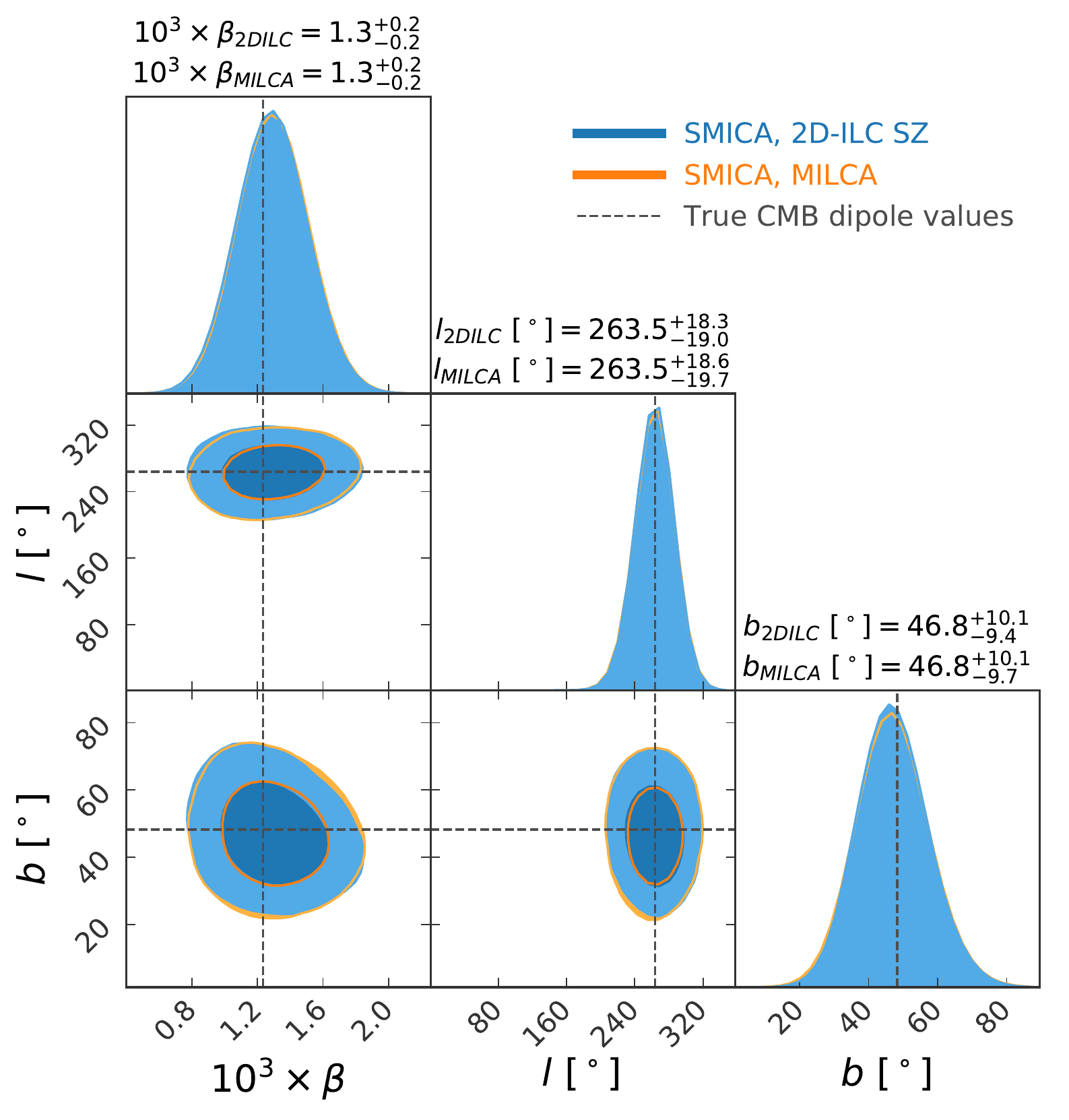}}
\mbox{\includegraphics[width=0.5\hsize]{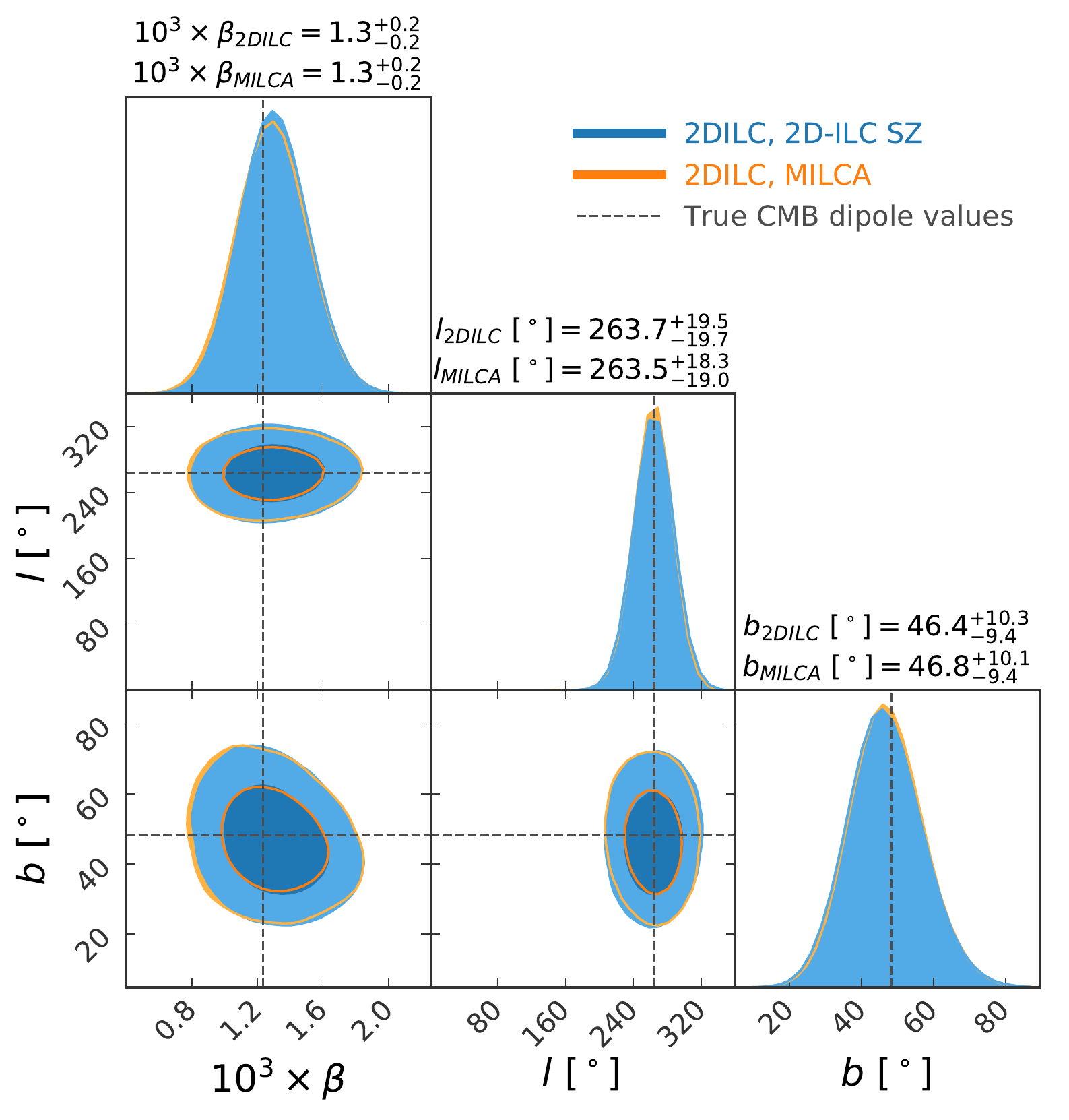}}
\caption{Posteriors for the CMB dipole parameters using the map-space analysis (top) and harmonic-space analysis (bottom). The left panels use the \smicano\ CMB maps, whereas the right use the \twodilc\ CMB maps. \milca\ $y$ map results are shown in orange, and \twodilc\ $y$ map results are shown in blue. Black dashed lines show the best-fit parameters from direct measurements of the CMB dipole.  Dark and light contours enclose 68\,\% and 95\,\%, respectively. Titles for each panel give the best-fit results, along with the 68\,\% uncertainties.}
\label{fig:triangle}
\end{figure*}

The main results of this paper are presented in Tables~\ref{tab:chi2} and ~\ref{tab:chi22} and
\fig\ref{fig:triangle}. They show how consistent the data are with the
presence (or non-presence) of the dipole term, and the recovered posteriors of the dipole parameters, respectively. In the following subsections we describe our results for each method in more detail.

\subsection{Map-space results}

\begin{figure*}[htbp!]
\mbox{\includegraphics[width=0.5\hsize]{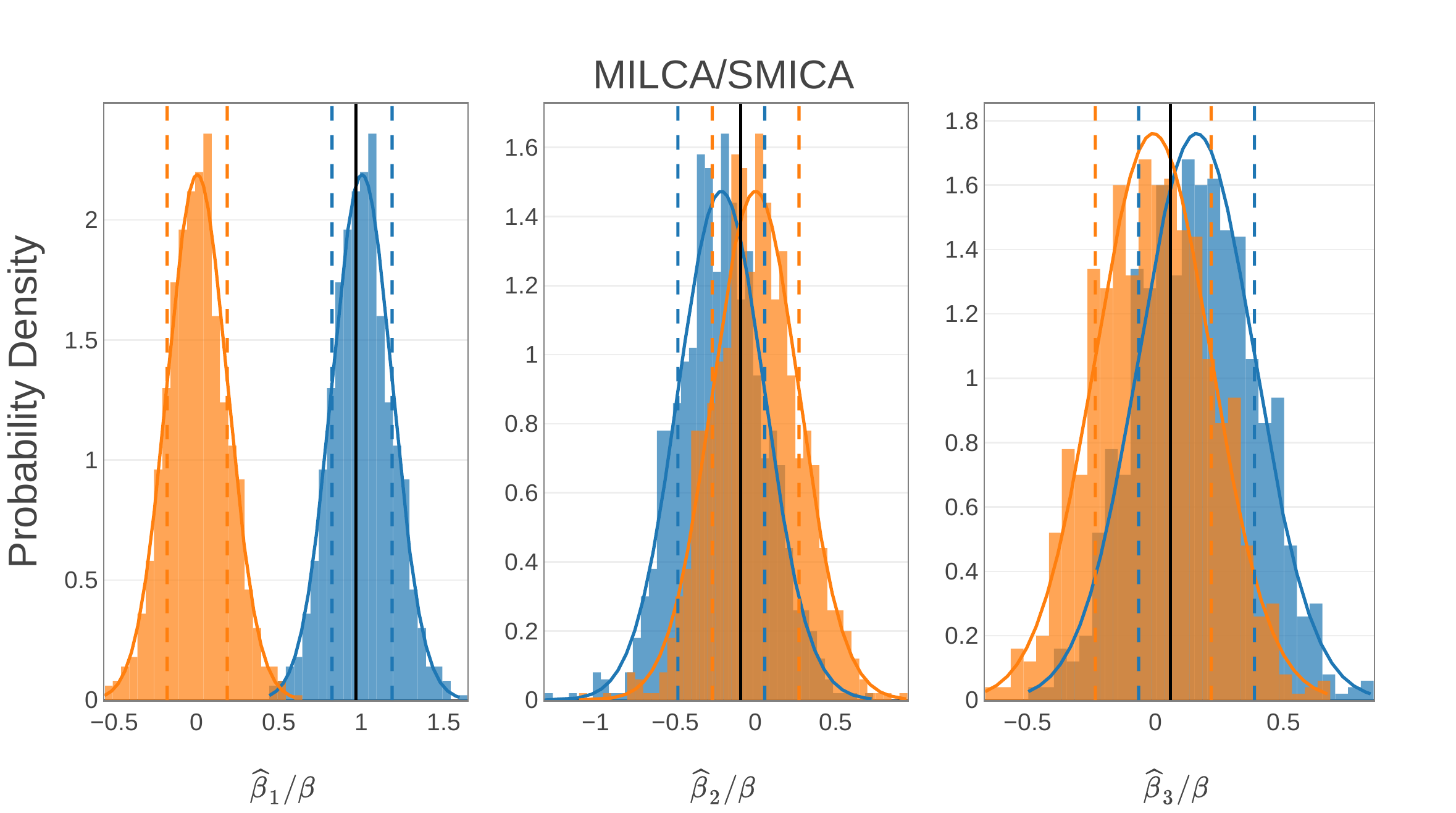}}
\mbox{\includegraphics[width=0.5\hsize]{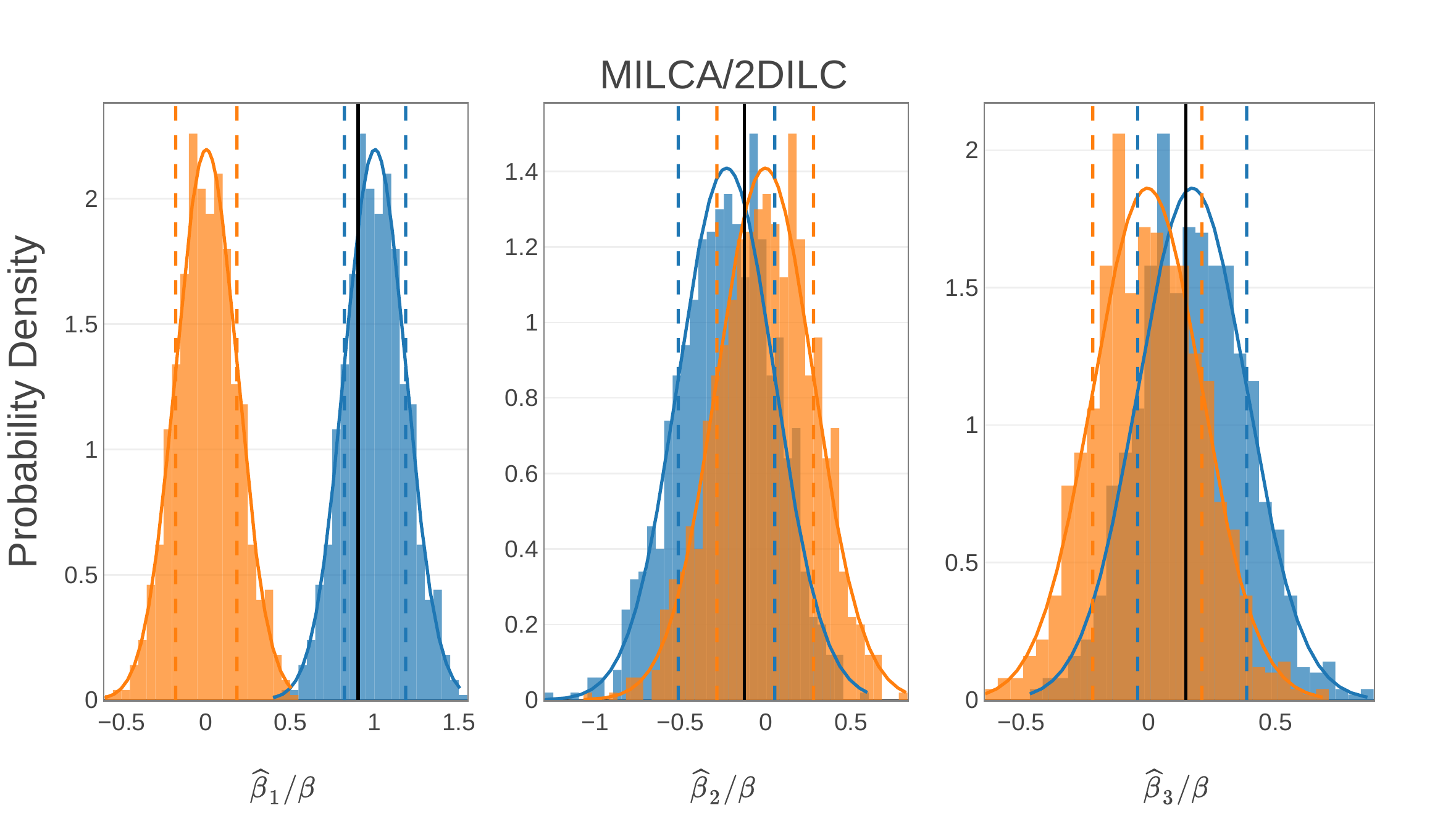}}
\mbox{\includegraphics[width=0.5\hsize]{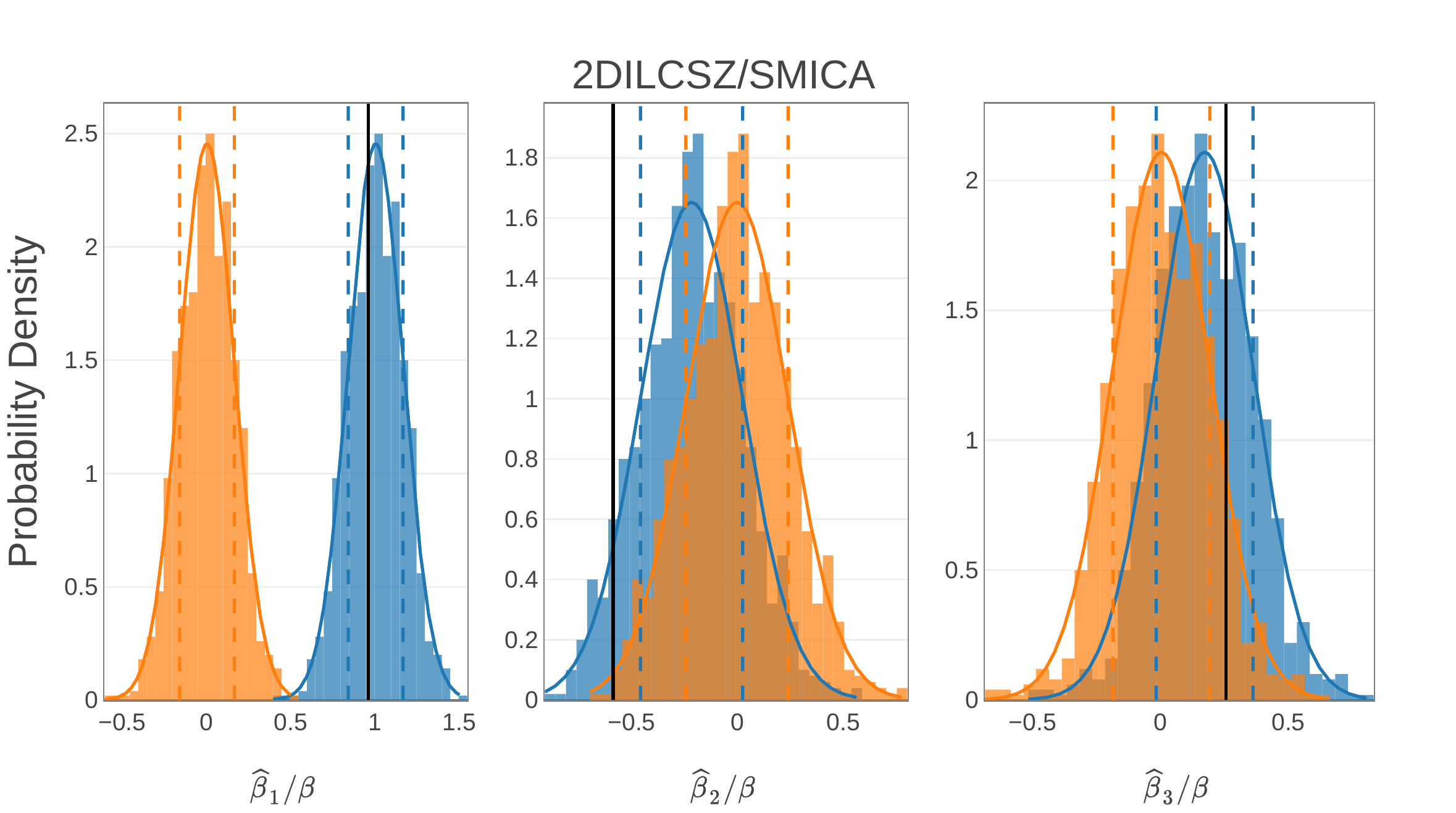}}
\mbox{\includegraphics[width=0.5\hsize]{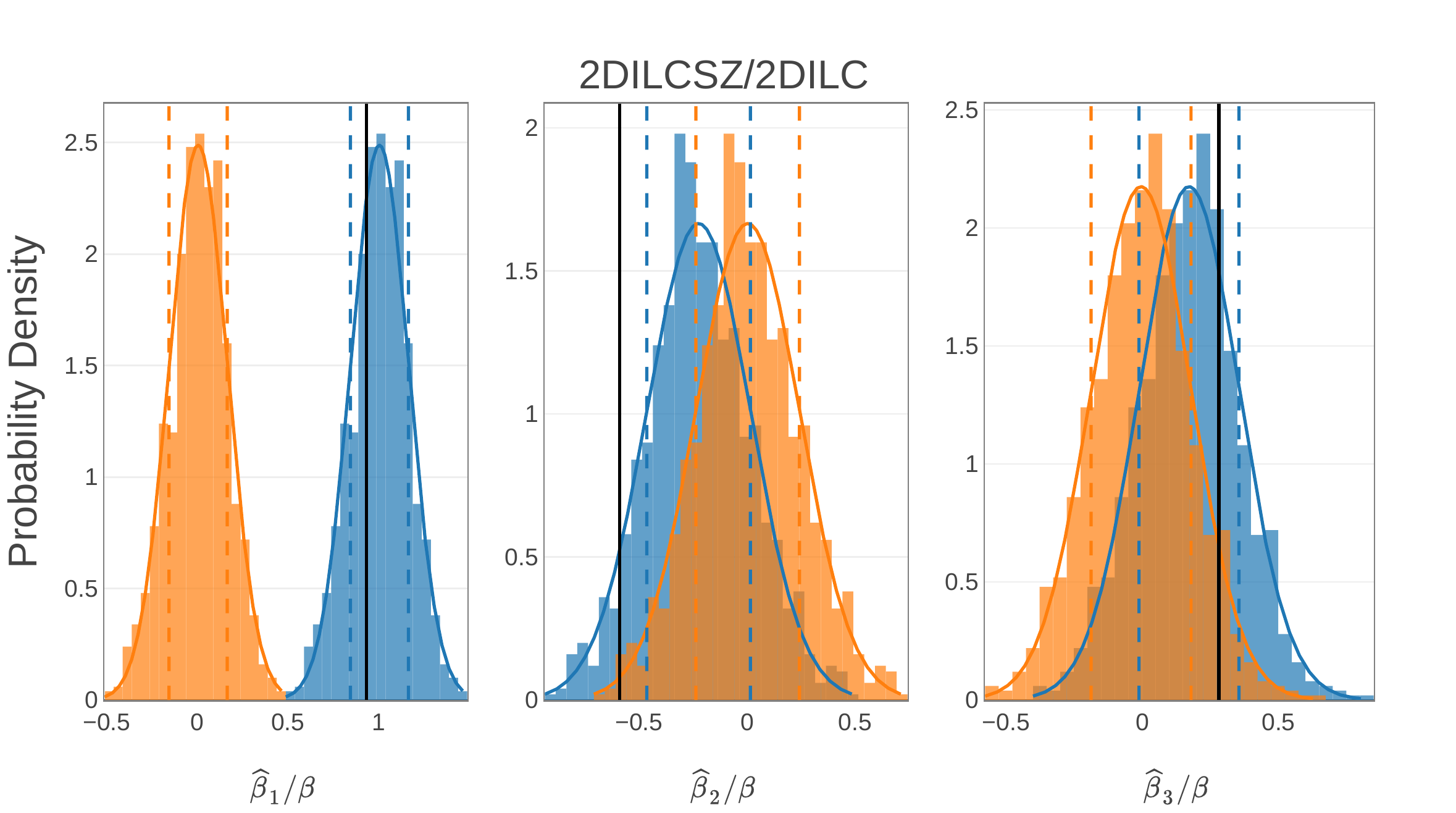}}
\caption{Histograms of $\hat{\beta}_i/\beta$ values (with 1, 2, and 3
corresponding to the CMB dipole direction, Galactic plane, and a third
orthogonal direction) using the map-space analysis for
\milca\ (top) and \twodilc\ (bottom) $y$~maps, and for CMB template maps \smicano\ (left) and \twodilc\ (right). Blue histograms are simulations with the dipolar modulation
term, and orange histograms are simulations without. Black vertical lines
denote the values of the data, demonstrating that they are much more consistent
with the existence of the dipolar modulation term than without it. Dashed lines show the 68\,\% regions for a Gaussian fit to the histograms. }
\label{fig:map_histo}
\end{figure*}

First we compare the consistency of the data with our two sets of simulations
(with and without the dipole term). This comparison shown in
\fig\ref{fig:map_histo}, with blue histograms being the simulations \emph{with} the dipole term and orange histograms \emph{without}. The data (black line) for \twodilc\
and \milca\ can clearly be seen to be consistent with the
simulations with the dipole term; this observation is made quantitative from
examination of the $\chi^2$ (see Tables~\ref{tab:chi2} and \ref{tab:chi22}). The map-space method is more susceptible to biases induced by the mask, particularly in the off-dipole directions, $\hat{\beta_2}$ and $\hat{\beta_3}$; this is due to subtle correlations between the mask and templates, but has only a small effect in those directions
(at the level of a few tenths of $\sigma$), as can be seen in Fig.~\ref{fig:map_histo}. 
Converted into the equivalent probabilities for Gaussian statistics, we can
say that the dipole modulation is detected at the 5.0 to $5.9\,\sigma$ level.

\subsection{Harmonic-space results}

\begin{figure*}[htbp!]
\mbox{\includegraphics[width=0.5\hsize]{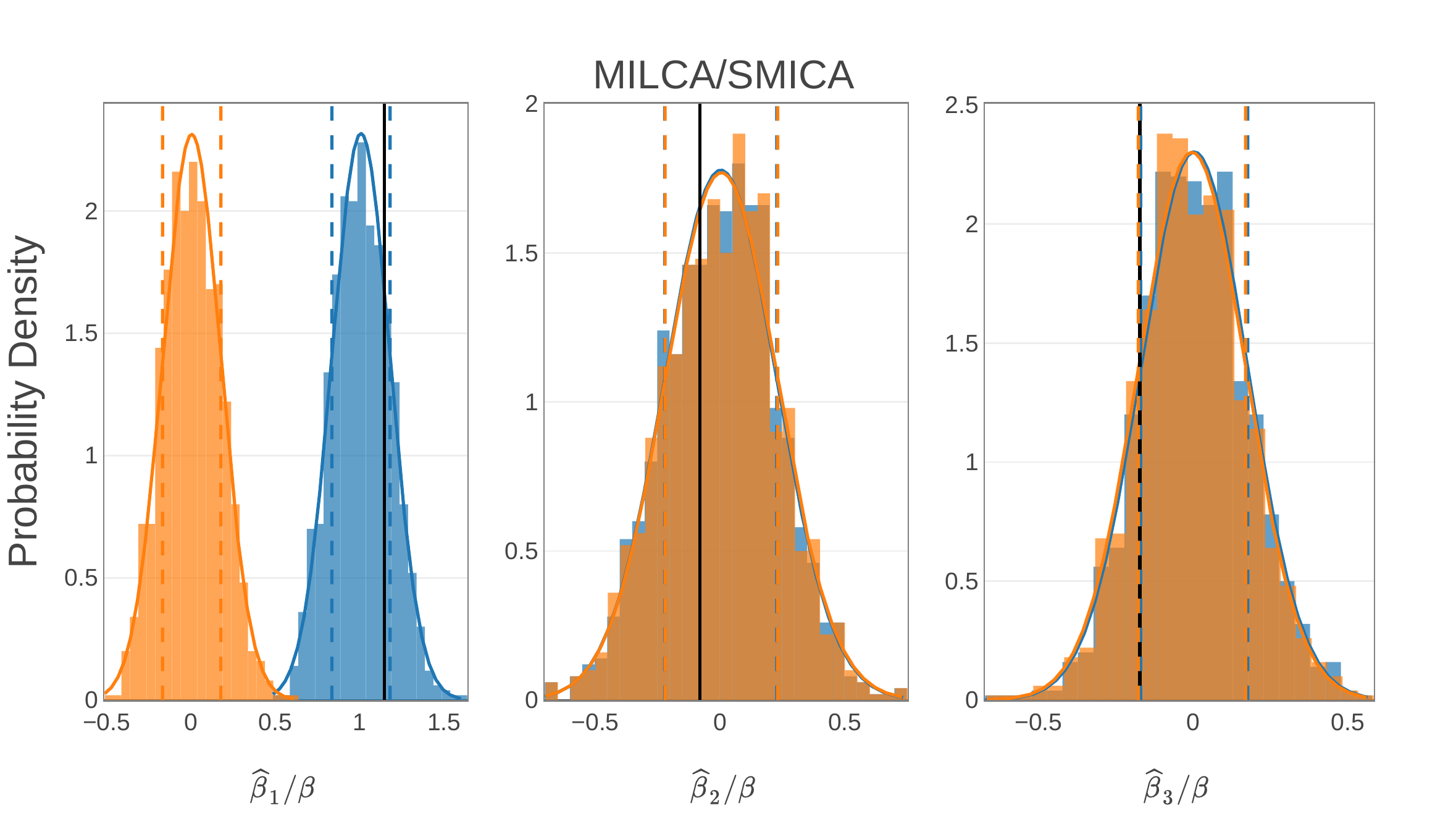}}
\mbox{\includegraphics[width=0.5\hsize]{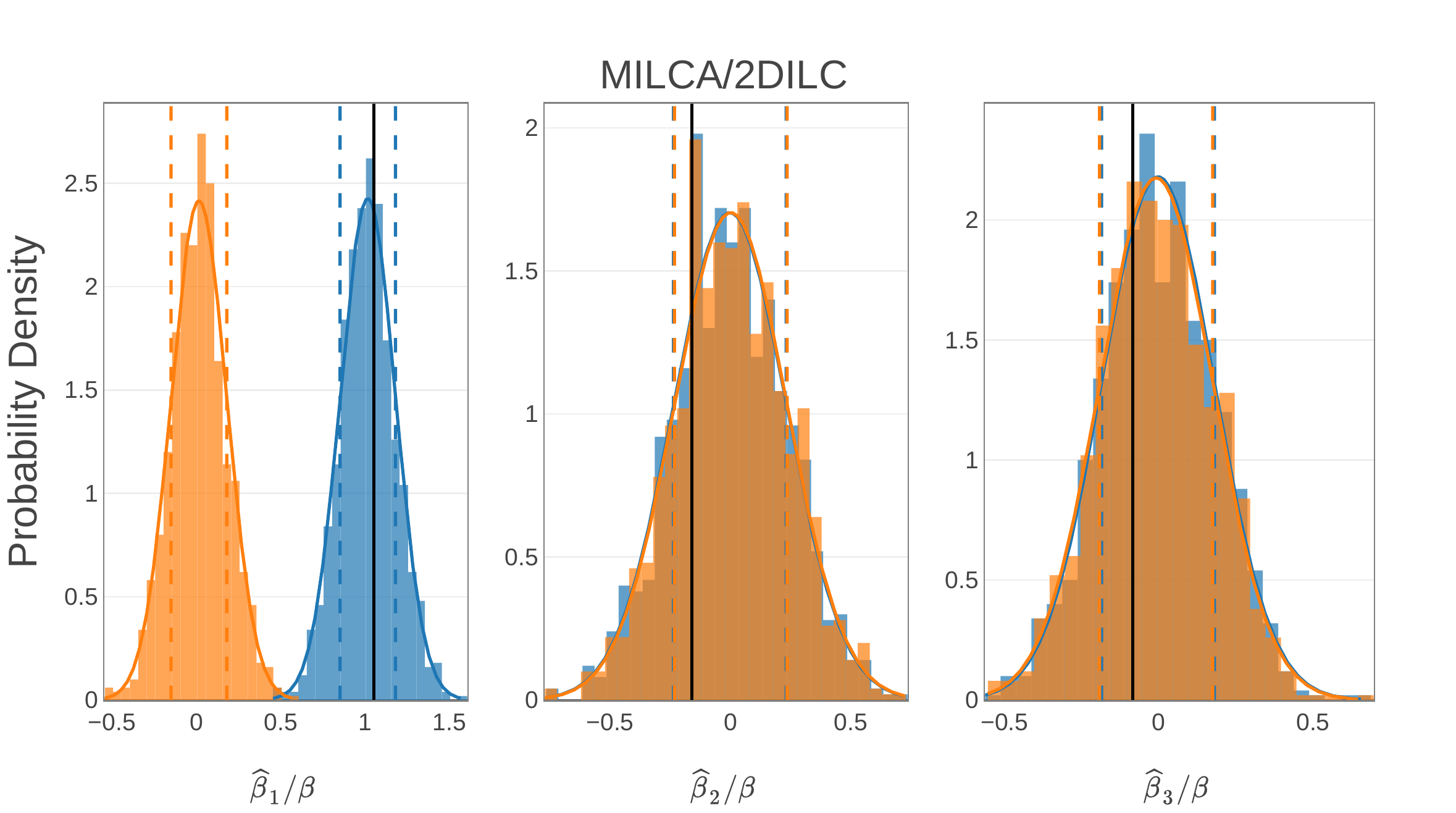}}
\mbox{\includegraphics[width=0.5\hsize]{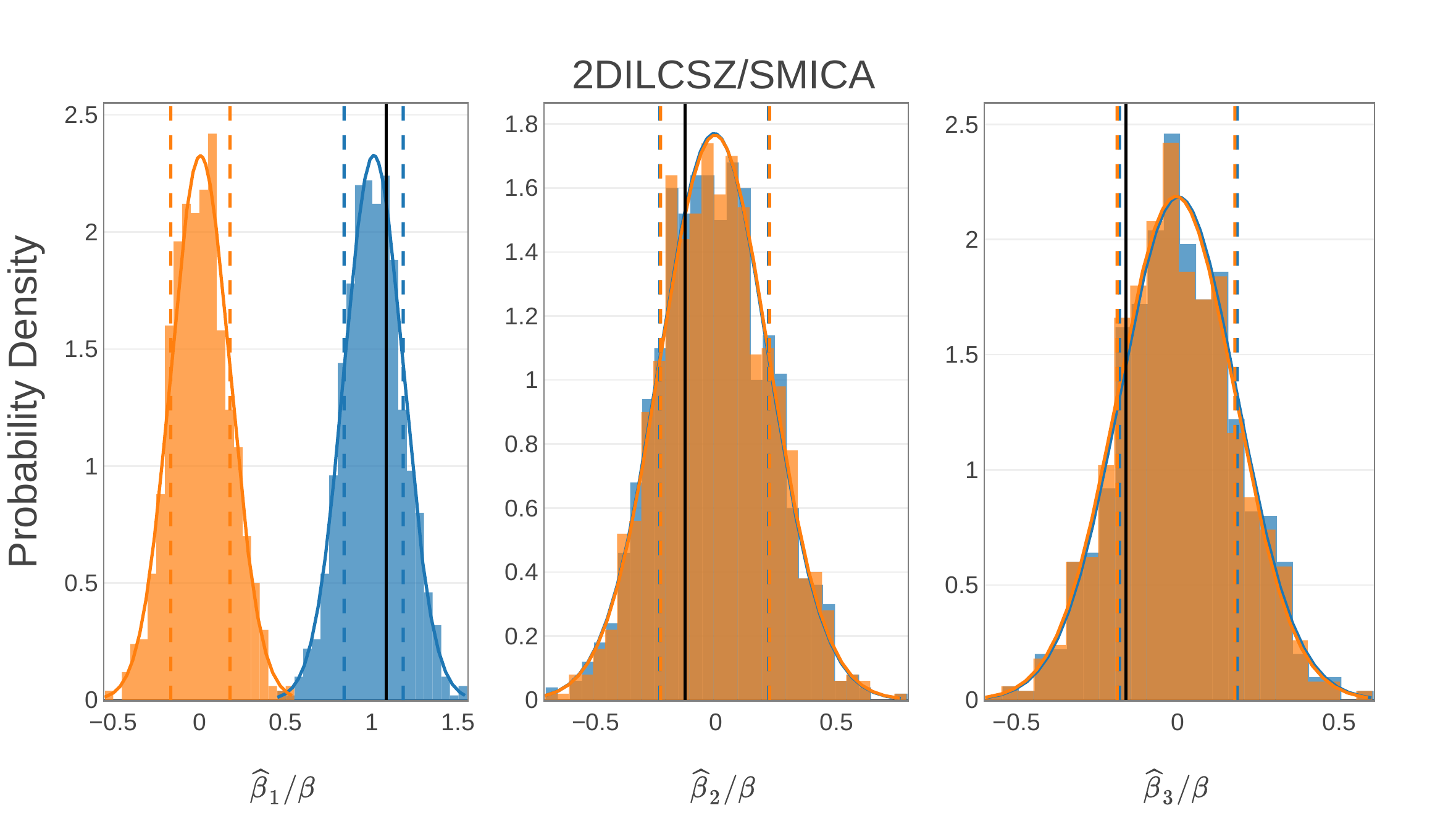}}
\mbox{\includegraphics[width=0.5\hsize]{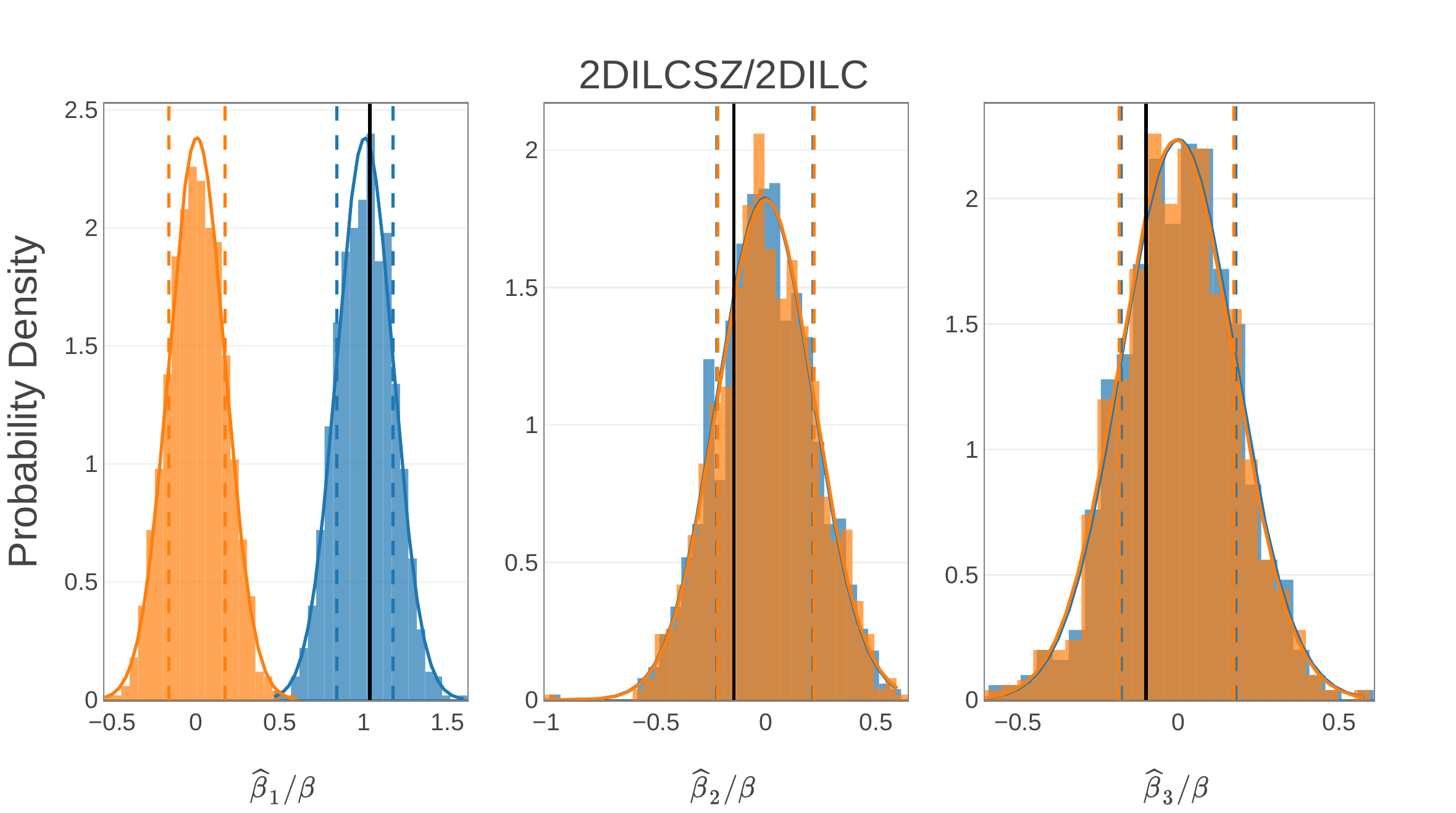}}
\caption{As in Fig.~\ref{fig:map_histo}, except now for the harmonic-space analysis.}
\label{fig:har_histo}
\end{figure*}

Figure~\ref{fig:har_histo} is the equivalent of \fig\ref{fig:map_histo}, but
for the harmonic-space analysis. Similar to the previous
subsection the data are much more consistent with the modulated simulations
than the unmodulated simulations. Tables~\ref{tab:chi2} and ~\ref{tab:chi22} contain the explicit
$\chi^2$ values and verify this quantitatively.
The harmonic-space method is somewhat susceptible to biases induced by the mask, due to the complex coupling that occurs, mainly between the $\ell$ and $\ell\pm2$ modes. This can be seen in the slight bias in the results for $\hat{\beta}_2$ and $\hat{\beta}_3$.
Nevertheless, we can say that we confidently detect the dipole modulation at
the 6.2 to $6.6\,\sigma$ level.

\section{Systematics}
\label{sec:systematics}
We have generated results using two distinct methods, namely the map-space method and harmonic-space method, with two distinct CMB maps and two distinct $y$~maps, and have shown the results to be consistent with the presence of a dipole-modulation signal in the expected direction. Each test is subject to slightly different systematics, but since the results are consistent, we can conclude that there is likely no significant systematic interfering with the results. Further tests, relaxing the limits of $\ell_{\rm max}=1411$ show that it is possible to achieve even higher levels of significance using smaller-scale data (see
Appendix~\ref{app:higherL}).  In that sense, the results in this paper are
conservative; however, if it becomes possible to construct reliable $y$~maps
out to higher multipoles then it should be possible to achieve a detection of
the dipole modulation at perhaps twice the number of $\sigma$ as found here.

\subsection{Residuals in the component separation}
\label{sec:residuals}
The \nilc\ $y$~maps are known to contain some remnant CMB contamination, unlike the \milca\ and \twodilc\ $y$~maps, which have been generated with the express purpose of eliminating the CMB contribution. This contaminates the signal we are looking for; the results from the \nilc\ $y$~maps may be seen in Appendix~\ref{app:nilc}. Any contamination remaining in the \milca\ and \twodilc\ $y$~maps is sufficiently low that is does not hide the dipole modulation signal.

\subsection{Galactic foregrounds}
It is known that the $y$~maps are contaminated by Galactic foregrounds; however, as the results here are from a cross-correlation of the modulated CMB maps with the $y$~maps such contamination does not have a large effect on the results. To further support this, a number of different mask sizes and combinations were tested, 
with the final mask selected because among those choices consistent with the more conservative masks, it gave the highest signal-to-noise ratio. Larger masks serve only to decrease the signal-to-noise of the data. This suggests that foregrounds have only a small effect on the detection of the dipole modulation. Foregrounds have been mentioned as a potential issue in previous results \citep{planck2014-a28}.

\section{Conclusions}
\label{sec:conclusions}
Due to the existence of the CMB dipole, a tSZ map necessarily contains a contaminating signal that is
simply the dipole modulation of the CMB anisotropies. This occurs because CMB
experiments do not directly measure temperature anisotropies, but instead measure
intensity variations that are conventionally converted to temperature variations. This
contamination adds power to the tSZ map in a $Y_{20}$ pattern,
with its axis parallel to the dipole direction. We have measured this effect
and determined a statistically independent value of the CMB dipole, which is
consistent with direct measurements of the dipole. Using a conservative multipole cut on the $y$~map, the significance of the detection of the dipole modulation signal is around 5 or $6\,\sigma$, depending on the precise choice of data set and analysis method. This is a significant improvement from the 2 to $3\,\sigma$ results in \citet{planck2013-pipaberration}. We also find that the contamination of the tSZ map contributes negligible noise to the bispectrum calculations (see Appendix~\ref{sec:bispectrum}).

\section*{Acknowledgements}
\label{sec:acknowledgements}

The \Planck\ Collaboration acknowledges the support of: ESA; CNES and CNRS/INSU-IN2P3-INP (France); ASI, CNR, and INAF (Italy); NASA and DoE (USA); STFC and UKSA (UK); CSIC, MINECO, JA, and RES (Spain); Tekes, AoF, and CSC (Finland); DLR and MPG (Germany); CSA (Canada); DTU Space (Denmark); SER/SSO (Switzerland); RCN (Norway); SFI (Ireland); FCT/MCTES (Portugal); and ERC and PRACE (EU). A description of the Planck Collaboration and a list of its members, indicating which technical or scientific activities they have been involved in, can be found at \href{http://www.cosmos.esa.int/web/planck/planck-collaboration}{\texttt{http://www.cosmos.esa.int/web/planck/}} \href{http://www.cosmos.esa.int/web/planck/planck-collaboration}{\texttt{planck-collaboration}}. Some of the results in this paper have been derived using the {\tt HEALPix} package and the {\tt NaMaster} package and some plots were generated using the {\tt pygtc} package.

\bibliographystyle{aat}
\bibliography{Planck_bib,ymodulation,software}

\appendix

\section{The tSZ bispectrum}
\label{sec:bispectrum}
Fundamentally the modulation is a correlation between $C_\ell$ and $C_{\ell\pm1}$. The signal considered here therefore shows up most prominently in the 4-point function (i.e., trispectrum) and thus we do not expect it to bias the measurements of the tSZ bispectrum; however, since the bispectrum is an important quantity for characterizing the tSZ signal, it is worth checking to ensure that the dipolar modulation does not add significant noise. In other words, we want to check if it is important to remove the dipole modulations before performing analysis of the tSZ bispectrum.
\citet{lacasa2012characterization} and \citet{bucher2010detecting} describe in detail the calculation of the bispectrum and the binned bispectrum, and this is summarized below.
The reduced bispectrum is given by
\begin{align}
  B_{\ell_1 \ell_2 \ell_3} &= (N_{\ell_1 \ell_2 \ell_3})^{-1/2} \notag \\
  & \times \sum_{m_1 m_2 m_3}
  \left(
  \begin{array}[]{ccc}
    \ell_1 & \ell_2 & \ell_3 \\
    m_1    & m_2    & m_3
  \end{array} \right)
  a_{\ell_1 m_1} a_{\ell_2 m_2} a_{\ell_3 m_3},
  \label{eq:bispectrum}
\end{align}
where $ \left( \begin{array}[]{ccc}
    \ell_1 & \ell_2 & \ell_3 \\
    m_1    & m_2    & m_3
\end{array} \right)$ represent the Wigner-3$j$ functions and \begin{align}
  N_{\ell_1 \ell_2 \ell_3}=\frac{(2\ell_1+1)(2\ell_2+1)(2\ell_3+1)}{4\pi}\left(\begin{array}[]{ccc}
  \ell_1 & \ell_2 & \ell_3 \\
  0      & 0      & 0
\end{array} \right)^2.
  \label{eq:bisNbin}
\end{align}
The normalized bispectrum is non-zero for terms where $m_1+m_2+m_3=0$, $|\ell_2-\ell_2|\ge \ell_3 \ge \ell_1+\ell_2$, and $\ell_1+\ell_2+\ell_3$ is even (this is due to the term before the sum with $m_1,m_2,m_3=0$). Typically, the binned bispectrum is analysed to reduce the number of terms calculated and saved, which constitutes only a small loss of information because the bispectrum is expected to vary slowly with $\ell$ \citep{lacasa2012characterization}. The data are binned by breaking down the interval from $\ell_{\rm min}$ to $\ell_{\rm max}$ into $i$ bins, denoted by $\Delta_i$. An average for the bispectrum of a particular bin can then be calculated using

\begin{align}
    B_{i_1 i_2 i_3}&=\frac{1}{\Xi_{i_1 i_2 i_3}}\sum_{\ell_1 \in \Delta_1}\sum_{\ell_2 \in \Delta_2}\sum_{\ell_3 \in \Delta_3} B_{\ell_1 \ell_2 \ell_3},
    \label{eq:bisBin}
\end{align}
where $\Xi_{i_1 i_2 i_3}$ is the number of non-zero elements in the given bin. Both the bispectrum and the binned bispectrum may be calculated using an integral over the map space as well, rather than in harmonic space.
This is achieved by first generating the binned scalemaps defined by
\begin{align}
  y_{\Delta_i}(\vec{\hat{n}})=\sum_{\ell \in \Delta_i, m} y_{\ell m} Y_{\ell m}(\vec{\hat{n}}),
  \label{eq:scalemaps}
\end{align}
where the sum goes from $\ell_{min}$ to $\ell_{max}$ in the bin $\Delta_i$. We can then use
\begin{align}
	B_{i_1 i_2 i_3}=\frac{1}{N_{i_1 i_2 i_3}}\int \mathrm{d}^2\vec{\hat{n}}\,y_{\Delta_1}(\vec{\hat{n}})\,y_{\Delta_2}(\vec{\hat{n}})\,y_{\Delta_3}(\vec{\hat{n}})
	\label{eq:bisBin-scale}
\end{align}
which gives the weighted average of the bispectrum within the bins \citep{lacasa2012characterization}.

\begin{figure*}[htbp!]
\mbox{\includegraphics[scale=1]{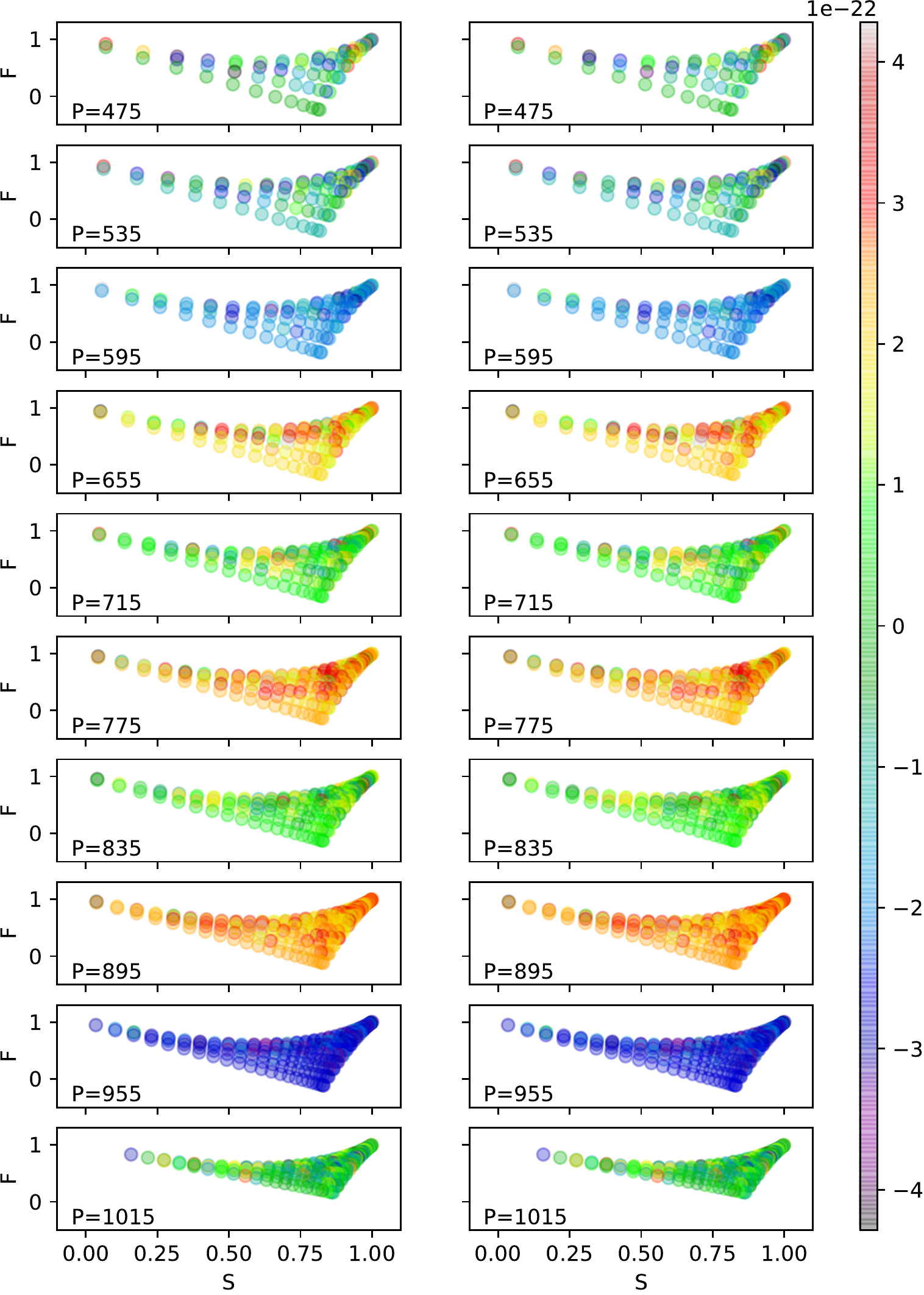}}
\caption{Binned bispectrum, for simulated $y$~maps, with $\ell_{\rm max}=500$ and bin sizes of 10, and ``scalemaps'' defined in Eq.~\ref{eq:scalemaps}, of $N_{\rm side}=512$. The left panels show the bispectrum for a simulated $y$~map {\it with\/} the dipole modulation and the right panels show the same with {\it no\/} dipole-modulation. The quantities $P$, $F$, and $S$ are as defined in Eqs.~\eqref{eq:bidefs}.}
\label{fig:BispectrumVal}
\end{figure*}

\begin{figure*}[htbp!]
\mbox{\includegraphics[scale=1]{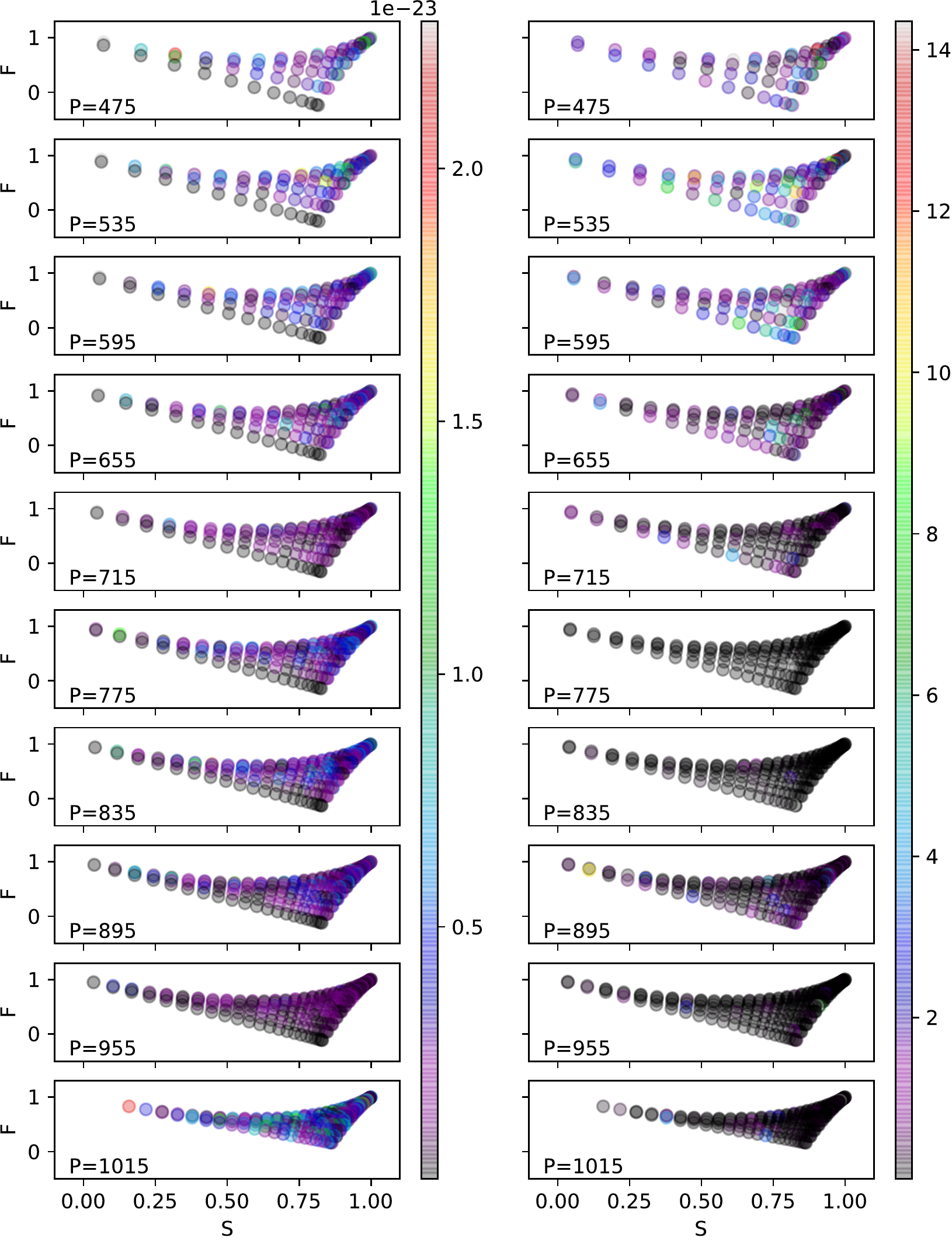}}
\caption{Absolute (left) and relative (right) difference between the bispectrum with and without the dipole-modulation term.}
\label{fig:BispectrumVal_diff}
\end{figure*}

In Figs.~\ref{fig:BispectrumVal} and \ref{fig:BispectrumVal_diff} we show a subset of the binned normalized bispectra for the $y$~maps, with and without the dipole modulation. For simplicity, since we are just comparing the results of two simulated maps, there are no non-Gaussianities and no mask applied. This analysis was performed using the \milca\ $y$~map and the \smicano\ CMB temperature map. Plots are constructed in the style suggested by \citet{lacasa2012characterization} for an $\ell_{\rm max}$ of 500 (and an \nside\ of 512 to speed up computation).

Useful definitions here are
\begin{align}
    \sigma_1&=\ell_1+\ell_2+\ell_3,\\
    \sigma_2&=\ell_1\ell_2+\ell_1\ell_3+\ell_2\ell_3,\\
    \sigma_3&=\ell_1\ell_2\ell_3,\\
    \Tilde{\sigma}_2&=12\sigma_2/\sigma_1^2-3,\ \in [0,1],\\
    \Tilde{\sigma_3}&=27\sigma_3/\sigma_1^3,\ \in [0,1],\\
    F&=32(\Tilde{\sigma_2}-\Tilde{\sigma_3})/3+1,\\
    S&=\Tilde{\sigma_3},\\
    P&=\sigma_1,
    \label{eq:bidefs}
\end{align}
where $P$ is the perimeter, each plot represents the results of a particular perimeter size, $F$ is plotted along the $y$-axis of the panels and $S$ is plotted along the $x$-axis of the panels.

Our main goal is to determine whether the dipole modulation contamination of the $y$~maps is significant, and to what degree it is significant for current and future analysis as data improves. For this purpose a subset of the tested perimeter values are plotted, for data with the dipole modulation and without, and the absolute value of the differences. It does not appear that the dipole modulation has a noticeable effect on the bispectrum results.

\section{NILC \textit{y}-map results}
\label{app:nilc}
\begin{figure*}[htbp!]
\mbox{\includegraphics[width=0.5\hsize]{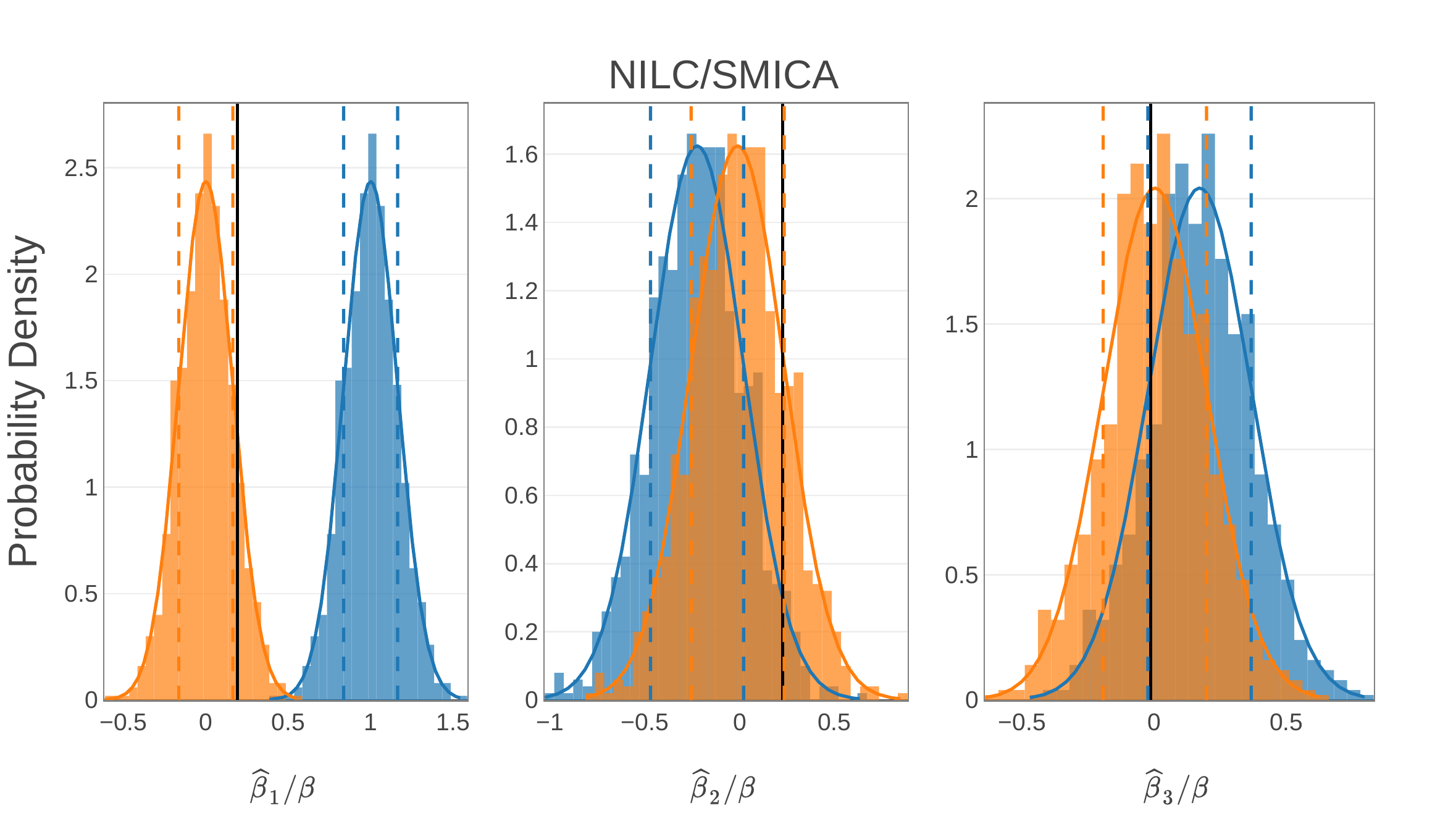}}
\mbox{\includegraphics[width=0.5\hsize]{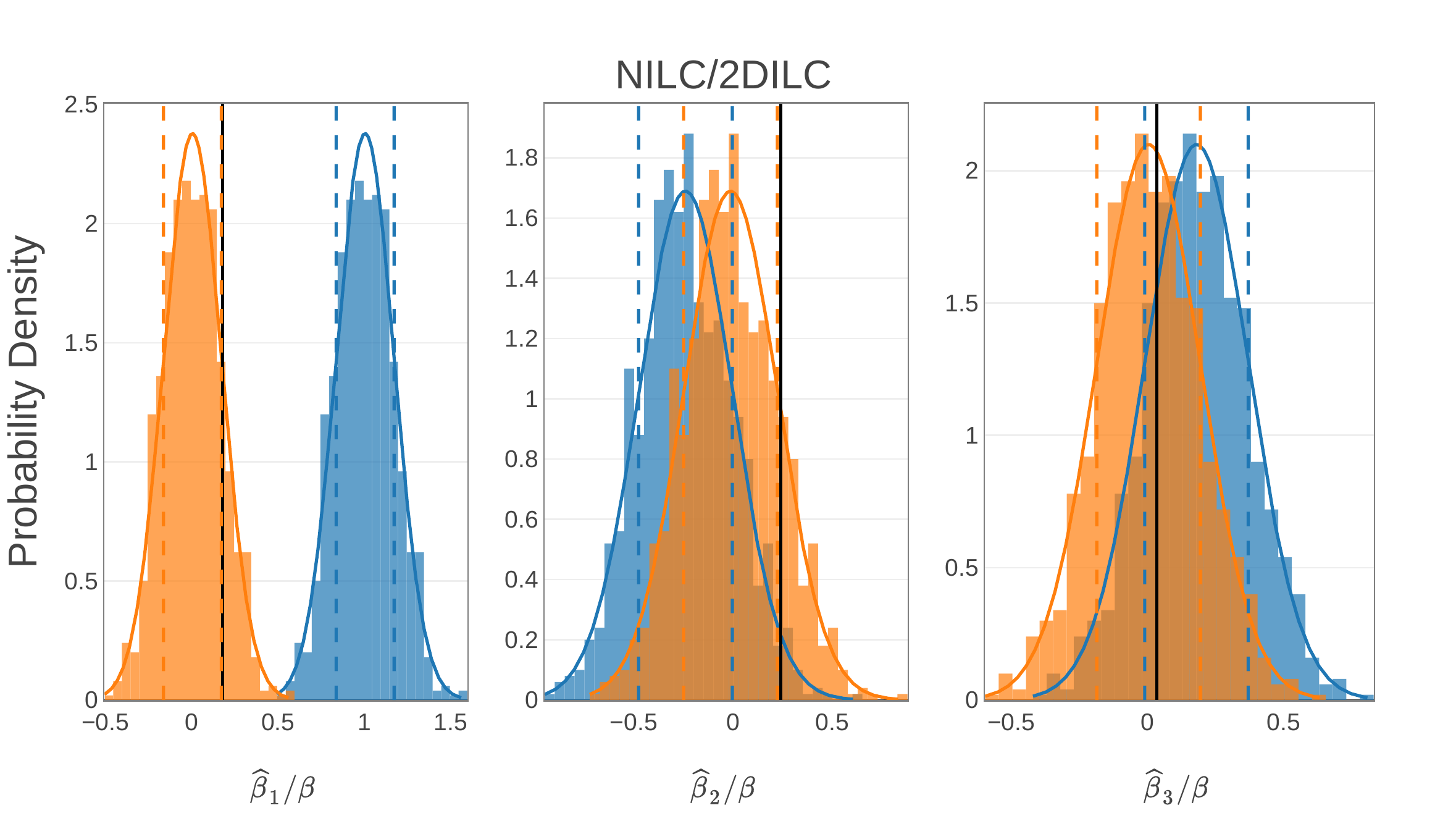}}
\mbox{\includegraphics[width=0.5\hsize]{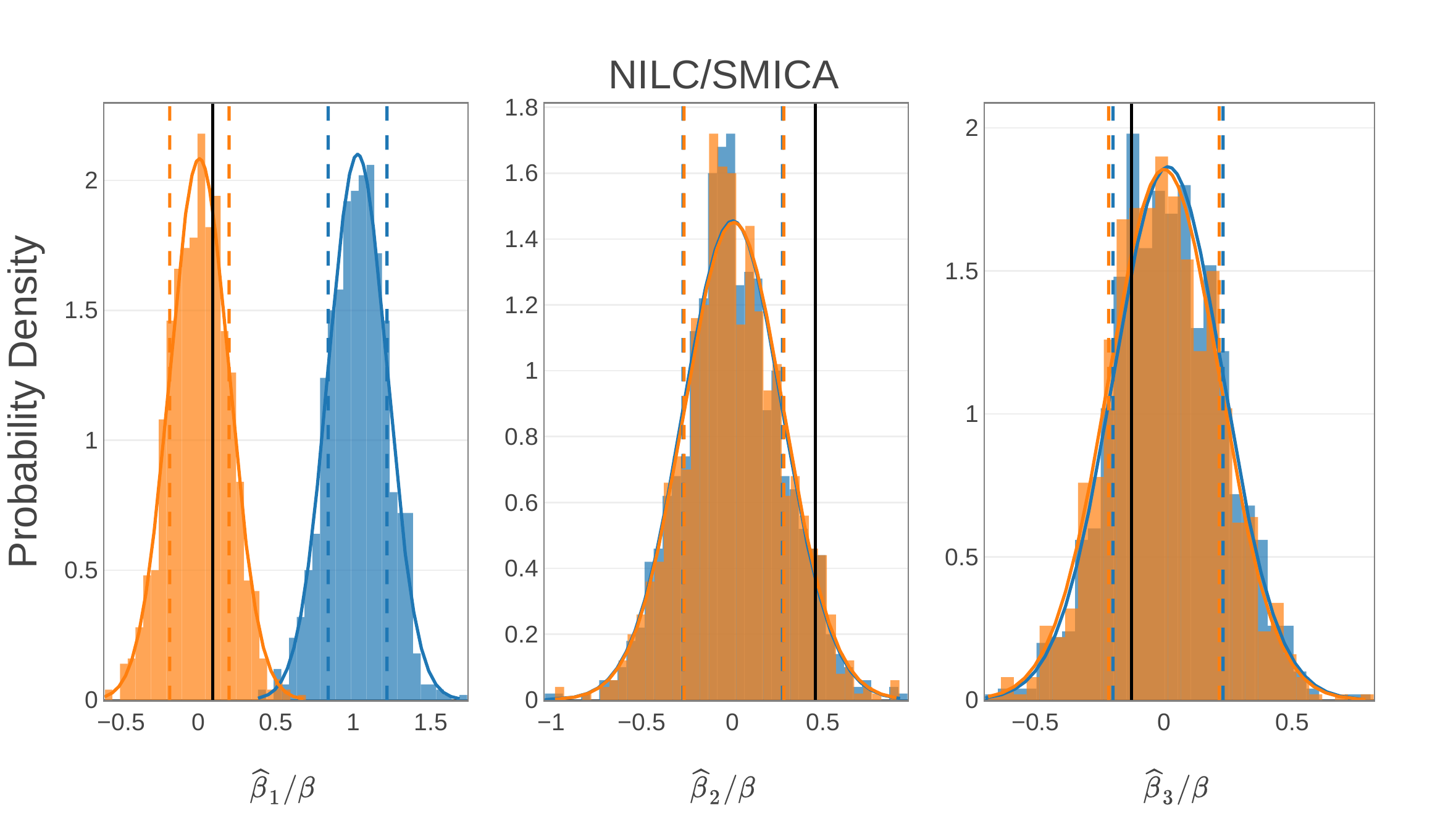}}
\mbox{\includegraphics[width=0.5\hsize]{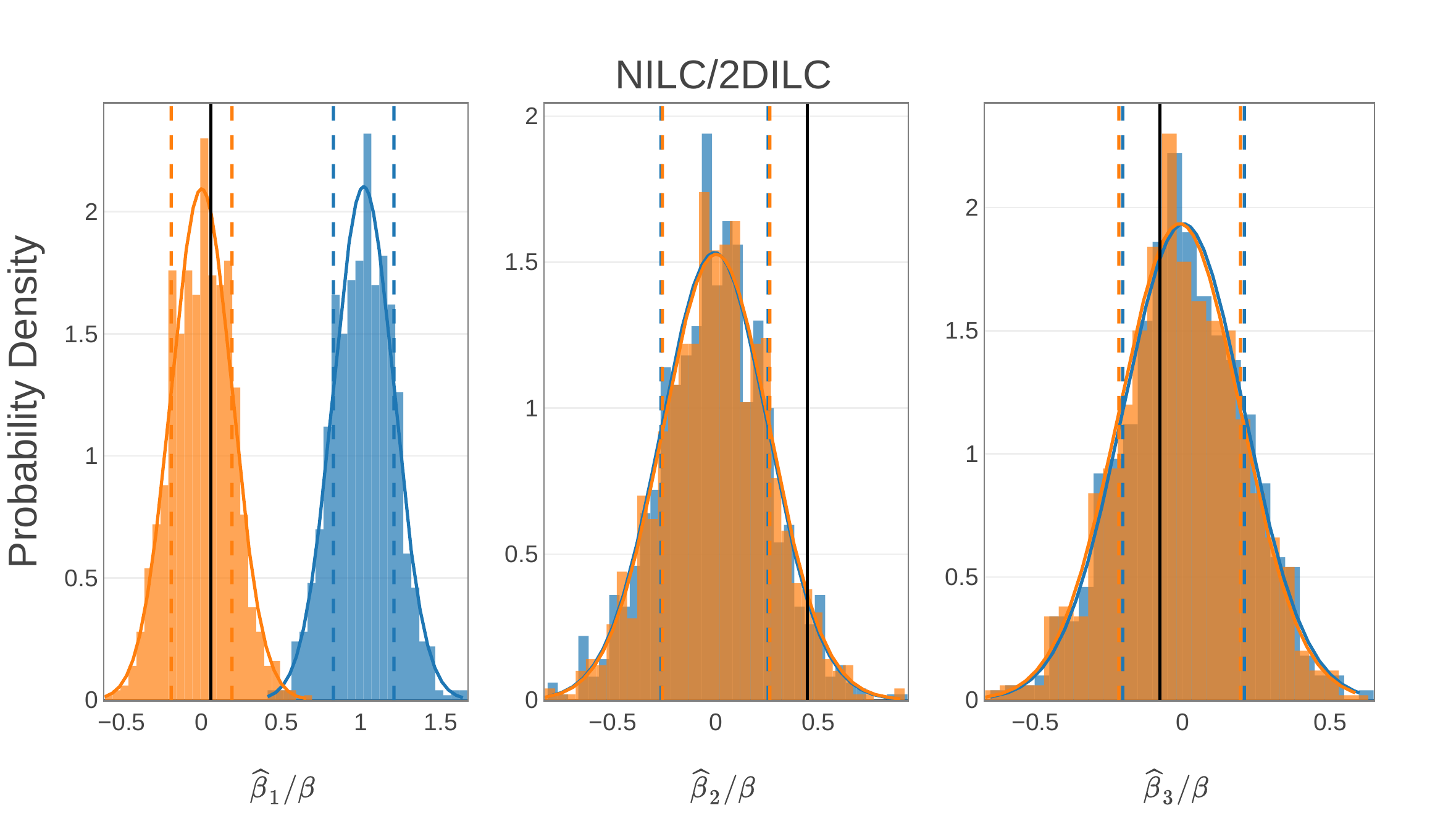}}
\caption{As in Figs.~\ref{fig:map_histo} (top) and \ref{fig:har_histo} (bottom), except using the \nilc\ $y$~maps. The top panels in this case show the results from the map-space method, while the bottom panels show those from the harmonic-space method.}
\label{fig:nilc_histo}
\end{figure*}

In creating the \Planck\ $y$~maps using \nilc, the choices were optimized for removal of the contamination by CMB, foregrounds, and noise. With the \milca\ $y$~maps there was an additional constraint added to fully eliminate the CMB, at the expense of adding more foregrounds and noise contamination. For this reason the CMB contamination in the \nilc\ $y$~maps is too high for us to robustly detect the dipole modulation. The \twodilc $y$~map was also produced with the express intent of removing all CMB contamination, and both it and the \milca\ maps clearly show that the dipole modulation is present. For completeness, here we present the effect of the contamination in the \nilc\ $y$~maps in Fig.~\ref{fig:nilc_histo}.  The dipole modulation signal is seen to be completely hidden by the CMB contamination.

\section{Increased \boldmath$\ell$\unboldmath$_{\bf max}$ results}
\label{app:higherL}
In our analysis for the harmonic-space method the results were truncated at $\ell_{\rm max}=1411$, since this is the recommendation from \cite{planck2014-a28} to avoid the correlated noise and foreground contaminations present in higher $\ell$. If we were to assume that the simulations model the data properly up to a higher $\ell_{\rm max}$, and that we also trust the data up to this higher $\ell_{\rm max}$, then we would be able to achieve a greater significance than reported in the conclusions. This can be seen in the simulation results using the \milca\ $y$~map and the \smicano\ CMB templates. These particular results are from 500 simulations for $N_{\rm side}=1024$, and $\ell_{\rm max}=2750$.  The significance appears to be at the ${>}\,12\,\sigma$
level. To do the analysis fully at this $\ell_{\rm max}$ the Weiner filter would also need to be applied to the CMB maps, as without it the bias would be much larger than the 2\,\% found in our analysis.

\begin{figure}[htbp!]
\mbox{\includegraphics[width=\hsize]{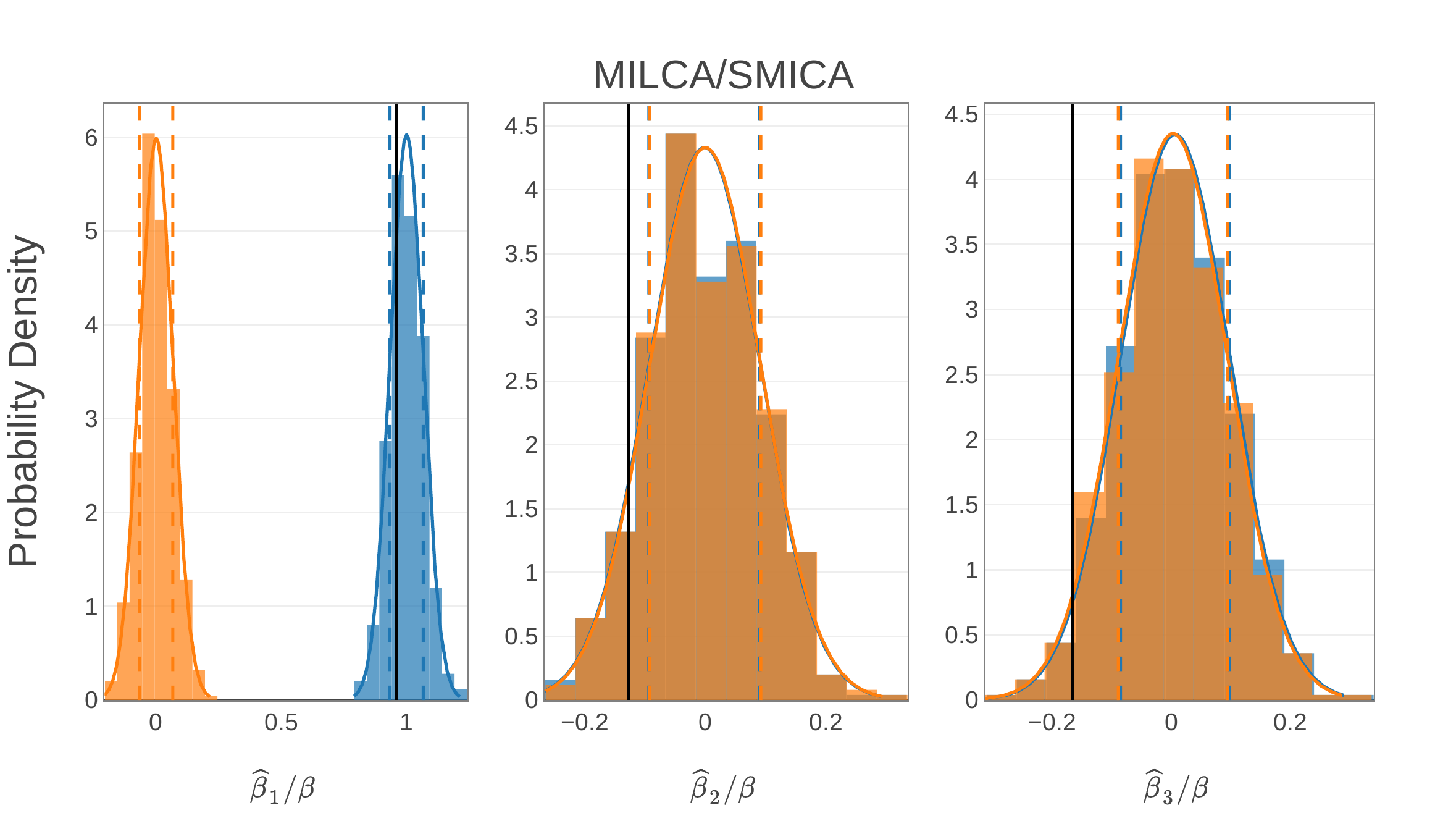}}
\caption{As in Fig.~\ref{fig:har_histo}, but with $\ell_{\rm max}=2750$ compared
to $\ell_{\rm max}=1411$ used in the paper. If we were to trust the $y$ map out to these multipoles, then these results would have a significance of ${>}\,12\,\sigma$.}

\label{fig:highl_histo}
\end{figure}

\raggedright

\end{document}